# Coarse-Grained Description of Stochastic Neutron Transport in Highly Heterogeneous Reactor Media via Non-Local Feller Boundary Functionals

V. V. Ryazanov, https://orcid.org/0000-0002-5308-3212

Institute for Nuclear Research, pr. Nauki, 47 Kiev, Ukraine, e-mail: vryazan19@gmail.com

This paper presents a novel hierarchical coarse-graining framework to model non-Markovian neutron transport within the ultra-heterogeneous, multi-phase subcooled nucleate boiling layer of a nuclear reactor core. By applying Feller's subordination theory to continuous-time random walks, microscopic fluid fluctuations and vapor bubble dynamics are rigorously integrated into macroscopic transport operators through the apparatus of Complete Bernstein Functions (CBF). A modified non-local Fick's law is derived using a $P_1$-Legendre polynomial expansion, mapping the fractal porosity and structural phase boundaries into a stochastically renormalized effective diffusion coefficient. The thermodynamic consistency of the model is grounded within Chentsov's statistical manifolds, demonstrating that granular Poisson-like shot noise yields a geometric-exponential family of distributions characterized by a constant negative information curvature. To evaluate operational safety margins under transient conditions, a generalized Lyapunov functional with a quadratic memory kernel is constructed, and a complex decomposition of the transfer function $H(\omega)$ is performed. The resulting analytical limits are parameterized for a WWER-1000 core layout and translated into an engineering computational sequence suitable for implementation in Mathcad Prime 8 to diagnose the onset of Hopf bifurcations and spatial power oscillations.

## 1. Introduction

The accurate modeling of neutron transport and thermal-hydraulic coupling under transient and accident conditions (Duderstadt and Hamilton, 1976; Lewis and Miller, 1984; D'Auria, 2017; Faucher, 2019) remains a cornerstone of nuclear reactor safety analysis. In high-power density systems, such as pressurized water reactors (PWR) and specifically WWER-1000, multi-phase flow regimes drastically alter the local moderating and absorbing properties of the coolant. A critical phenomenon in this context is subcooled nucleate boiling (Jens and Lottes, 1951; Rohsenow, 1952; Chen, 1966; Wallis, 1969; Collier and Thome, 1994; RELAP5 Development Team, 1995) (wall boiling) occurring on the fuel rod cladding surface under intense heat fluxes, even when the bulk coolant temperature remains well below saturation. The continuous generation, growth, and subsequent condensation of vapor micro-bubbles form a highly heterogeneous, stochastic boundary layer.

In conventional reactor physics codes (e.g., RELAP5 (RELAP5 Development Team, 1995) or traditional coupled CFD-neutron kinetics models (Pereira et al., 2007)), these multi-phase effects are typically resolved through static spatial homogenization and deterministic averaging based on the classical transport theory (Hébert, 1993; Yasseri and Rahnema, 2014). The underlying macroscopic governing equations, such as the standard Boltzmann equation or the classic neutron diffusion equation, implicitly rely on Markovian assumptions and Fick's law (Bell and Glasstone, 1970; Duderstadt and Hamilton, 1976). These formulations presume a separation of scales where the mean free path of a neutron significantly exceeds the characteristic correlation length of structural or fluid phase fluctuations. However, during rapid transients or localized power excursions, the boiling layer instantly transforms into an ultra-heterogeneous, polydisperse medium. Under these conditions, the classic assumption of a localized, memoryless transport operator breaks down (Pomraning, 1989; Behringer et al., 1979). Large vapor cavities induce anomalous spatial trajectories ("Lévy flights"), while bubble retention dynamics introduce long-range temporal correlations (non-Markovian dynamic memory) (Montroll and Scher, 1973; Corona, 2016). Neglecting these fluctuations can lead to dangerous overestimations of the departure from nucleate boiling (DNB) ratio or fail to predict localized spatial power oscillations (Koreshi, 2002).

To bridge the gap between microscopic stochastic fluctuations and macroscopic deterministic safety margins, the statistical physics concept of *coarse-graining* must be re-evaluated (Van Kampen, 1991; Prasad and Rahnema, 2008). Traditional coarse-graining techniques focus on static spatial cell homogenization, which merely renormalizes effective macroscopic cross-sections but preserves the parabolic nature of the transport operators (Larsen, 1975). Recent mathematical advancements in stochastic thermodynamics and Feller's subordination theory (Feller, 1971) demonstrate that macroscopic time evolution can be rigorously defined via boundary functionals (e.g., first-passage time, local time) of the underlying microscopic trajectories (Ryazanov, 2026a, 2026b). When integrated into information geometry, specifically within the framework of Chentsov's statistical manifolds (Chentsov, 1982; Amari, 2016), the transformation from a micro-description to a macro-description imposes strict algebraic stability constraints on the choice of the temporal subordinator. For systems dominated by granular, Poisson-like (shot noise) fluctuations at the phase boundaries — which are intrinsically characteristic of nuclear reactor kinetics — the mapping yields a specific geometric-exponential family of distributions characterized by a constant negative information curvature (Amari, 1985; Ruppeiner, 1995).

Building upon these foundational principles, this paper presents a novel hierarchical coarse-graining model for neutronic transport in stochastic media. The micro-scale fluid fluctuations within the subcooled boiling layer are treated as continuous time random walks (CTRW) subordinated by a dynamic boundary functional. By utilizing the apparatus of Complete Bernstein Functions (CBF) with a truncated power-law spectral measure (Metzler and Klafter, 2000; Schilling et al., 2012), we explicitly incorporate the fractal porosity of the boiling layer along with the hydrodynamic bubble lifetime dictated by the modified Labuntsov-Kolykhan heat transfer relations (Labuntsov, 1984; Kolykhan and Pulyayev, 1987).

The primary objectives of this study are twofold:

1. To rigorously derive a modified non-local Fick's law from the generalized transport equation and evaluate how the uncoupled log-periodic and power-law spectral measures modify the effective diffusion length and homogenization limits.
2. To formulate the macroscopic reactor kinetics equations under the influence of this memory kernel and analyze the stability boundaries via explicit Lyapunov functionals and complex transfer function mappings.

Ultimately, this approach provides a robust analytical link between microscopic multi-phase fluid dynamics and macroscopic operational safety parameters, ensuring that high-fidelity simulations remain free from computational singularities near bifurcation thresholds.

Building upon this hierarchical framework, the present study investigates the coupled frequency-domain and time-domain stability of a closed-loop WWER-1000 reactor system governed by neutron kinetics and thermal-hydraulic void feedback under ultra-heterogeneous core conditions. During startup, low-power operational modes, or localized bulk subcooling fluctuations, non-Gaussian statistics and stochastic clustering of the neutron population can induce significant reactivity noise. By scaling the heterogeneous medium through a logarithmic subordinator, the phenomenological transport coefficients are explicitly linked to the microscopic grain scale $d_{cell}$, the clustering of nucleation sites r, and the characteristic vapor bubble lifetime $\tau_c$. The transition from non-local memory effects to localized Markovian transport is controlled by the spatial kernel truncation parameter $\gamma$.

To guarantee that inherent fluctuations are efficiently suppressed by internal feedbacks, this work bridges the gap between frequency-domain stability criteria and explicit time-domain trajectories. Within the frequency domain, the stability margin is evaluated via the Nyquist and Mikhailov criteria by analyzing the characteristic matrix and mapping the critical coolant bulk subcooling threshold $\Delta T_{\mathbf{sub,crit}}$. It is demonstrated that crossing this boundary triggers a Poincaré–Andronov–Hopf bifurcation, transforming a stable phase-portrait focus into a limit cycle associated with sustained density wave oscillations. Furthermore, parametric sensitivity analysis reveals how variations in the linear heat generation rate q (which governs the static heat transfer timescale) and the critical feedback frequency $\omega_c$ expands the auto-oscillatory domain, demanding higher operational subcooling. Finally, to validate these frequency-domain margins in the time domain, a discrete recursive numerical scheme is implemented to track the relaxation of the core temperature T(t) following an initial thermal perturbation, confirming a smooth, monotonically damped return to steady-state equilibrium within less than 0.8 seconds.

The article is organized as follows. Section 2 presents the mathematical framework, covering coarse-graining and Feller subordination. Section 3 examines the physical mechanisms of grain formation in nuclear reactors. Section 4 derives the modified Fick's law from the non-local transport equation. Section 5 addresses thermal-hydraulic coupling, specifically Labuntsov bubble hydrodynamics. Section 6 discusses the thermodynamic Lyapunov functional and non-local stability. Section 7 provides an analytical decomposition of the complex transfer function H(ω). Section 8 covers the stability of phase portraits and bifurcation analysis. Section 9 presents the results and discussion, and Section 10 offers conclusions.

## 2. Mathematical Framework: Coarse-Graining and Feller Subordination

### 2.1. Coarse-Graining Procedure and Entropy Redefinition

To transition from a microscopic description of a highly heterogeneous medium to a macroscopic transport model, we implement a statistical projection technique. Let Γ denote the microscopic phase space of the system, and let $\rho(q,p,t)$ be the exact microscopic probability density function satisfying the reversible Liouville equation. We partition Γ into a finite set of macroscopic volumetric cells (grains) $\{\Delta\Gamma_k\}$, where k is the discrete cell index.

The macroscopic observer is unable to distinguish between distinct microstates residing within the same cell. Consequently, we define the coarse-grained (averaged) probability density $\bar{\rho}_k$ (t) as:

$$\bar{\rho}_k(t) = \frac{1}{\Delta\Gamma_k}\int_{\Delta\Gamma_k} \rho(q,p,t)\,d\Gamma .$$

While the microscopic Gibbs entropy remains constant under reversible Hamiltonian evolution $dS_{Gibbs}/dt=0$, the systematic loss of micro-information inherent in the coarse-graining procedure breaks time-reversal symmetry. The macroscopic entropy production is evaluated by monitoring the evolution of the coarse-grained density $\bar{\rho}_k$.

To encompass non-Markovian memory effects induced by trapping regimes at phase boundaries, we generalize the classical Boltzmann-Gibbs framework. When the microscopic fluctuations transition from standard exponential families to heavy-tailed or long-range correlated distributions, the macro-entropy yields a non-linear operator structure related to the geometric-exponential family. For a macroscopic probability distribution P(x), the generalized coarse-grained entropy $S_{macro}[P]$ is defined via the inverse of the governing Bernstein function f(s) (Appendix A.1) as:

$$S_{macro}[P] = -k_B\int_{\Omega} P(x)\cdot f^{-1}(\ln P(x))\,dx\,, \tag{1}$$

where $k_B$ is the Boltzmann constant, and Ω represents the spatial domain of the macro-system. Under this formulation, the rate of entropy production is explicitly governed by the spectral characteristics of the grain boundaries.

### 2.2. Feller Subordination via Boundary Functionals

Following the semi-group approach pioneered by Feller (1971), the dynamic evolution of a stochastic process in a complex medium can be modeled by decoupling the physical time t from an intrinsic, operational macro-time τ. Let Y(t) represent a microscopic transport process (e.g., localized neutron scattering or velocity fluctuations). We introduce a non-decreasing stochastic subordinator τ(t), which physically represents the cumulative operational time spent by the particle interacting with the internal boundaries of the grain.

The physical mechanism behind this subordination is governed by a continuous additive boundary functional $A_t$ (such as the local time (Ryazanov, 2025) or absorption functional on the grain surface). The operational macro-time process τ(t) is formally defined as the right-continuous inverse of the boundary functional $A_t$, mapping directly to a First-Passage Time (FPT) problem:

$$\tau(t) = \inf\{s > 0 : A_s > t\} . \tag{2}$$

The subordinated macroscopic process Z(t) is then constructed via the time-deformation mapping:

$$Z(t) = Y(\tau(t)) . \tag{3}$$

In the language of characteristic functions (CF), if the unperturbed micro-process has a cumulant (logarithm of the CF) denoted by $\Psi_{cell}(\beta,\lambda)$ , the subordinated macro-state is determined by the composition with a Bernstein function f(s). The generalized macroscopic partition function (or characteristic operational functional) $Z_{macro}$ is expressed as:

$$Z_{macro}[\beta,\lambda,\mu] = \exp(-\int_0^\infty f(\Psi_{cell}(\beta,\lambda))w(d\lambda)) , \tag{4}$$

where $w(d\lambda)$ is the macroscopic structural scaling measure (Appendix A.2).

### 2.3. Complete Bernstein Functions and Spectral Measures

To ensure that the coarse-graining mapping remains physically consistent and does not violate the causality or infinite divisibility of the underlying semi-groups, Chentsov's closedness condition requires $f(s)$ to belong to the class of Complete Bernstein Functions (CBF). A function f:(0, ∞)→(0, ∞) is a CBF if it admits the standard integral Lévy-Khinchin (Pick-Nevanlinna) representation (Schilling et al., 2012):

$$f(s) = a + bs + \int_0^\infty (s / s + t)\mu(dt) , \tag{5}$$

where *a, b*≥0, and $\mu(dt)$ is a positive Borel spectral measure on (0, ∞) satisfying the regularizing condition $\int_0^\infty (1+t)^{-1}\mu(dt) < \infty$ .

The spectral measure $\mu(dt)$ represents the continuum of relaxation times (or trapping scales) dictated by the topological complexity of the grain boundaries. In this framework, we analyze two distinct structural limits:

Case A: Truncated Power-Law Spectral Measure (Polydisperse Medium).

For a highly disordered, multi-phase boundary layer—such as the subcooled nucleate boiling layer on a WWER fuel rod—the distribution of vapor bubble scales exhibits fractal characteristics with a sharp cutoff driven by turbulent shear fluid forces. The corresponding spectral measure is defined as:

$$\mu(dt) = (\sin(\alpha\pi) / \pi)t^{\alpha-1}e^{-\gamma t}dt , \tag{6}$$

where $0 < \alpha < 1$ represents the fractal indexing parameter of the porous topology, and γ>0 is the exponential regularization parameter. Substituting this measure into the CBF integral representation yields the macro-scale cumulant deformation, which transitions from fractional subdiffusion at short scales $t \ll \gamma^{-1}$ to classic Markovian diffusion at large scales ($t \gg \gamma^{-1}$).

Case B: Log-Periodic Oscillating Measure (Discrete Hierarchical Medium)

For strictly hierarchical, self-similar porous networks (e.g., regular fractal trees or fixed scaling lattices with a discrete scaling ratio $\kappa > 1$), the spectral measure concentrates into discrete lines:

$$\mu(dt) = \sum_{n=-\infty}^{+\infty} C_n \delta(t - \kappa^n)dt , \tag{7}$$

where $\delta$ is the Dirac delta function, and $C_n = \kappa^{-n\alpha}$ represents the scaling weight. In the time domain, this measure generates a memory kernel modulated by log-periodic oscillations (Appendix A.3):

$$M(t) = t^{-\alpha}[A_0 + \sum_{m=1}^{\infty} A_m \cos(2\pi m(\ln t / \ln\kappa) + \theta_m)] . \tag{8}$$

This structure induces periodic "bursts" of particle release from the boundary layers whenever the observation timescale matches a specific structural level of the coarse-grained hierarchy.

## 3. Physical Mechanisms of Grain Formation in Nuclear Reactors

Within the proposed coarse-graining framework, a physical "grain" represents a spatio-temporal cluster or a macroscopic element of the medium within which the nuclear-physical and thermodynamic properties of the fluid can be considered locally homogeneous. The boundary functionals governing the particle trajectories at the interfaces of these elements dictate the non-Markovian nature of macroscopic transport. The physical interpretation of these grains depends fundamentally on the reactor design.

### 3.1. Boiling Water Reactors and RBMK Systems

Under bulk two-phase boiling conditions, the structural grains are defined by the stochastic macroscopic phase distributions of the mixture:

- **Vapor Bubbles and Cavities:** These form geometric grains of extremely low density where the macroscopic cross-sections approach zero, leading to vacuum-like particle streaming ("streaming paths").
- **Liquid Bridges and Slugs:** Dense clusters of the liquid phase (water) where intense neutron moderation and absorption take place. The chord lengths and volumetric distributions of these grains follow a polydisperse allocation, which is analytically modeled via the fractal index $\alpha$.

### 3.2. Molten Salt Reactors (MSR)

In liquid-fueled systems where the fissile material is dissolved directly within circulating fluoride or chloride molten salts, the coarse-grained domains are governed by the hydrodynamic structure of the flow:

- **Turbulent Macro-Eddies (Fluid Parcels):** Strong thermal gradients and non-uniform power fields induce the formation of large eddy zones with varying local density and localized concentrations of fissile isotopes. Each turbulent parcel behaves as an individual grain possessing distinct homogenized cross-sections $\Sigma_f$ and $\Sigma_s$.
- **Helium Bubbles:** Gas bubbles injected to strip volatile fission products (such as xenon-135) act as microscopic voided grains of varying absorption capacity.

### 3.3. Pressurized Water Reactors (WWER)

Despite the system pressure suppressing bulk boiling, intense local thermal loads on the fuel rod cladding surfaces under nominal conditions induce specialized localized grain layers:

- **Subcooled Boiling Boundary Grains:** A thin boundary layer composed of fluctuating micro-bubbles that continuously generate and condense within the subcooled liquid bulk. The lifetime of these grains, $\tau_c$, is rigidly bounded by the condensation rate.
- **Turbulent Thermal Vortices:** Fluid parcels carrying localized temperature variations. Since water acts as the primary moderator, hot and cold fluid eddies form a stochastic field of macroscopic thermal moderation cross-sections.
- **Boric Acid Concentration Transients:** Microscopic regions experiencing localized fluctuations in dissolved $H_3BO_3$ during fast transient shifts, acting as highly absorbing ("black") or highly transparent ("white") clusters for the diffusing neutron field.

## 4. Derivation of the Modified Fick's Law from the Non-Local Transport Equation

To rigorously establish the macroscopic diffusion equation within the coarse-grained framework, we apply the spherical harmonics method ($P_N$ -approximation) to the non-local integro-differential transport equation. We restrict our derivation to the $P_1$ -approximation, assuming a weakly anisotropic neutron field within a highly scattering, polydisperse subcooled boiling layer.

The generalized one-dimensional, stationary non-local neutron transport equation for the angular flux density Ψ(x,μ) is formulated as:

$$\mu \frac{d\Psi(x,\mu)}{dx} + \int_{-\infty}^{x} K(x-x^{'}) \cdot \Psi(x^{'},\mu) dx^{'} = \frac{\Sigma_s}{2} \int_{-1}^{1} \Psi(x,\mu^{'}) d\mu^{'} + Q(x), \quad (9)$$

where μ is the cosine of the neutron flight angle relative to the spatial axis x, $\Sigma_s$ is the macroscopic scattering cross-section, $Q(x)$ is the independent isotropic source, and K(x-x\`) is the spatial non-local kernel generated by the inverse Feller subordinator at the grain boundaries.

### 4.1. Legendre Polynomial Expansion ($P_1$-Approximation)

We expand the angular flux density Ψ(x,μ) in (9) into a truncated series of Legendre polynomials $P_k(\mu)$, retaining only the first two moments (k=0 and k=1):

$$\Psi(x,\mu) \approx \Phi(x)/2 + 3\mu J(x)/2,$$

where the scalar neutron flux $\Phi(x)$ and the net neutron current J(x) represent the fundamental integral moments of the angular distribution:

$$\Phi(x) = \int_{-1}^{1} \Psi(x,\mu) d\mu,$$

$$J(x) = \int_{-1}^{1} \mu\Psi(x,\mu) d\mu.$$

Substituting the $P_1$ expansion into the non-local transport equation (9) yields:

$$\mu \frac{d}{dx}\left[\frac{1}{2}\Phi(x) + \frac{3}{2}\mu J(x)\right] + \int_{-\infty}^{x} K(x-x^{'}) \cdot \left[\frac{1}{2}\Phi(x^{'}) + \frac{3}{2}\mu J(x^{'})\right] dx^{'} = \frac{\Sigma_s}{2}\Phi(x) + Q(x). \quad (10)$$

### 4.2. System of Angular Moments Equations

To extract the coupled equations for Φ(x) and J(x), we perform angular integration. First, integrating the transport equation (10) with respect to μ over the interval [−1,1], and utilizing the mathematical orthogonality conditions $\int_{-1}^{1} \mu d\mu = 0$ and $\int_{-1}^{1} \mu^2 d\mu = 2/3$, we obtain the macroscopic continuity (balance) equation:

$$\frac{dJ(x)}{dx} + \int_{-\infty}^{x} K(x-x^{'})\Phi(x^{'}) dx^{'} = \Sigma_s \Phi(x) + 2Q(x). \quad (11)$$

Second, we multiply the expanded transport equation (10) by $\mu$ and integrate over the same limits [−1,1]. Using the relations $\int_{-1}^{1} \mu^3 d\mu = 0$ and $\int_{-1}^{1} \mu^4 d\mu = 2/5$, the higher-order terms vanish, leading to the dynamic equation for the neutron current:

$$\frac{1}{3}\frac{d\Phi(x)}{dx} + \int_{-\infty}^{x} K(x-x^{'}) J(x^{'}) dx^{'} = 0. \quad (12)$$

### 4.3. Localization of the Non-Local Spatial Memory Kernel

To transition to the macroscopic scale (x >> 1/γ), where the individual grain fluctuations undergo asymptotic homogenization, we introduce a coordinate transformation under the integral: y=x-x\`. We then expand the displaced neutron current function J(x-y) into a Taylor series in the vicinity of the macroscopic point x: $J(x-y) \approx J(x) - y\frac{dJ(x)}{dx} + \frac{y^2}{2!}\frac{d^2 J(x)}{dx^2} - \dots$ .

Substituting this Taylor expansion back into the integral term of the current equation (12) results in:

$$\frac{1}{3}\frac{d\Phi(x)}{dx} + J(x)\int_0^\infty K(y)dy - \frac{dJ(x)}{dx}\int_0^\infty yK(y)dy + \ldots = 0 .$$

For a polydisperse subcooled boiling boundary layer governed by a Complete Bernstein Function, the spatial tracking kernel is modeled via the truncated power-law density:

$$K(y) = y^{-\alpha} e^{-\gamma y} / \Gamma(1-\alpha) .$$

We evaluate the structural spatial moments of this kernel explicitly:

- The Zero-th Moment (**$\Sigma_{t,eff}$**)**:** Defines the effective macroscopic homogenization cross-section of the granular mixture:

$$\Sigma_{\mathbf{t,eff}} = \int_0^\infty K(y)dy = \gamma^{\alpha-1} .$$

- The First Moment: Defines the non-local spatial shift correction:

$$\int_0^\infty yK(y)dy = (1-\alpha)\gamma^{\alpha-2} .$$

Under the conditions of high condensation scaling within the WWER primary circuit, the regularizing truncation parameter is large ($\gamma >> 1$). Consequently, the first and all higher-order spatial moments decay as $O(\gamma^{\alpha-2})$, allowing us to safely truncate the expansion. The localized equation for the neutron current simplifies asymptotically to:

$$\frac{1}{3}\frac{d\Phi(x)}{dx} + \Sigma_{t,eff} J(x) = 0 . \qquad (13)$$

### 4.4. Formulation of the Modified Fick's Law

Isolating the net neutron current J(x) from the localized equation (13), we establish the direct analytical link between the flux gradient and particle current:

$$J(x) = -\frac{1}{3\Sigma_{t,eff}}\frac{d\Phi(x)}{dx} .$$

To relate the spatial attenuation parameter $\gamma$ to the kinetic characteristics of the medium, we match the temporal and spatial decay scales of the neutron flux in the linear transport approximation. The survival probability of a neutron against condensation or capture over a time interval t is governed by the exponential factor $\exp(-t/\tau_c)$ where $\tau_c$ is the characteristic condensation time. Within the trajectory functional formalism, the corresponding spatial statistical weight over a path length x is represented as $\exp(-\gamma x)$. Applying the kinematic relation $t = x / v_n$ for neutrons propagating with an average velocity $v_n$, we equate the exponents $\exp(-x / v_n \tau_c) = \exp(-\gamma x)$, which directly yields $\gamma = 1 / v_n \tau_c$.

By defining the effective coarse-grained diffusion coefficient as $D_{eff} = 1/3\Sigma_{t,eff}$ and substituting the structural expression for the zero-th moment through the physical parameter of bubble lifetime $\gamma=1/v_n\tau_c$, we arrive at the **Modified Fick's Law**:

$$J(x) = -D_{eff}\frac{d\Phi(x)}{dx} , \qquad (14)$$

where the non-local stochastic parameters ($\alpha$, $\tau_c$) of the subcooled boiling grains are fully mapped into the algebraic macro-coefficient:

$$D_{eff} = (v_n\tau_c)^{1-\alpha} / 3 . \qquad (15)$$

Substituting this modified current expression into the continuity equation closes the framework, transforming the non-local operator system into a classical parabolic differential equation with stochastically renormalized macroscopic constants, protecting numerical codes from structural singularities.

## 5. Thermal-Hydraulic Coupling: Labuntsov Bubble Hydrodynamics

To close the coarse-grained neutronic transport model, we establish a direct physical link between the macroscopic technological variables of the reactor core and the microscopic operational grain lifetime $\tau_c$. Under highly subcooled conditions with turbulent forced convection, the bubble dynamics are mapped using the modified Labuntsov-Kolykhan heat transfer relations.

### 5.1. Phase Transition Mass Balance

The complete physical duration of a vapor bubble on the fuel cladding surface includes the initial growth stage $\tau_g$ and the subsequent condensation phase $\tau_{cond}$ following detachment:

$$\tau_c = \tau_g + \tau_{cond}.$$

Under the high-velocity turbulent conditions typical of a WWER-1000 core, the detached bubbles spend the vast majority of their existence collapsing within the subcooled boundary layer, rendering the initial growth duration negligible ($\tau_g << \tau_{cond}$). Thus, the operational lifetime collapses to:

$$\tau_c \approx \tau_{cond}.$$

### 5.2. Labuntsov Phase Collapse Differential Equation

According to Labuntsov's formulation, the radial collapse rate of a spherical vapor bubble *R(t)* within a fluid under a localized subcooling margin $\Delta T_{sub}$ is dictated by transient thermal layer conduction:

$$\frac{dR(t)}{dt} = -\frac{Nu \cdot \lambda_f \cdot \Delta T_{sub}}{2 \cdot r \cdot \rho_v \cdot R(t)}, \tag{16}$$

where $\lambda_f$ is the thermal conductivity of the liquid water phase, $r$ is the latent heat of vaporization, $\rho_v$ is the saturated vapor density at core pressure (15.7 MPa); $\Delta T_{sub} = T_{sat} - T_f$ represents the localized bulk subcooling margin, with $T_{\mathbf{sat}}$ denoting the coolant saturation temperature at the operating pressure and $T_f$ denoting the local bulk fluid temperature.

### 5.3. Turbulent Stream Drag Corrections

For a bubble tracking through a turbulent bulk flow with a velocity vector w, the localized Nusselt number scales with the square root of the Reynolds number. In the linear approximation form established by Kolykhan and Pulyayev (1987), this scales as:

$$Nu = Nu_0 (1+Bw).$$

Integrating the radial differential relation (15) from the detachment envelope (t=0, R=$R_0$) to full collapse (t=$\tau_c$, R=0):

$$\int_{R_0}^{0} RdR = -\int_0^{\tau_c} \left(\frac{Nu_0(1+Bw)\lambda_f \Delta T_{sub}}{2r\rho_v}\right)dt,$$

$$-\frac{R_0^2}{2} = -\left(\frac{Nu_0(1+Bw)\lambda_f \Delta T_{sub}}{2r\rho_v}\right)\tau_c.$$

This yields the raw dependency of grain lifetime on local bulk subcooling and flow speed:

$$\tau_c = \frac{r\rho_v R_0^2}{Nu_0(1+Bw)\lambda_f \Delta T_{sub}}. \tag{17}$$

### 5.4. Heat Flux Self-Similarity Scaling and Kinetic Closure

### 5.4.1. Functional Analysis and Dynamic Memory Asymptotics

By invoking Jakob's scaling laws (Appendix A.4), the initial detachment radius $R_0$ of a vapor bubble generated within the subcooled boiling boundary layer scales directly with the operational linear heat flux q and inversely with the localized bulk subcooling margin $\Delta T_{sub}$:

$$R_0 \sim q / \Delta T_{sub}. \tag{18}$$

Substituting this boundary scale constraint into the quadratic bubble lifetime equation (17) derived in Section 5.3 completes the multi-phase thermal-hydraulic closure relation. This mathematical mapping shifts the microscopic framework to measurable engineering variables, providing the correct non-linear analytical representation for the dynamic subordinator scale $\tau_c$:

$$\tau_c = \frac{A \cdot q^2 \cdot r \cdot \rho_v}{\lambda_f \cdot \Delta T_{sub}^3 \cdot (1 + B \cdot w)}, \tag{19}$$

where the thermal-hydraulic variables are defined as follows: *A* is a non-dimensional structural proportionality constant accounting for the geometric form factor of the subcooled boiling layout, *q* is the localized linear heat flux from the fuel rod cladding $\mathrm{W/m^2}$, *r* is the latent heat of vaporization (specific enthalpy of phase transition) J/kg, $\rho_v$ is the saturated vapor density at core pressure kg/m$^3$, $\lambda_f$ is the thermal conductivity of the liquid water phase $\mathrm{W}/(m \times K)$, $\Delta T_{sub}$ is the localized bulk coolant subcooling margin K or $^\circ\mathrm{C}$, B is the empirical turbulent stream drag coefficient s/m, *w* is the bulk coolant flow velocity vector m/s.

Evaluating this expression using the nominal material properties of water at WWER-1000 operational envelopes yields the corrected functional engineering representation:

$$\tau_c = \frac{4.2 \cdot 10^{-3} \cdot q^2}{\Delta T_{sub}^3 \cdot (1 + 0.15 \cdot w)}. \tag{20}$$

The quantity *A*, along with the constant thermophysical properties of water at an operating pressure of 15.7 MPa $r, \rho_v, \lambda_f$ and the base Nusselt number $Nu_0$, has already been incorporated into a single numerical factor $4.2 \cdot 10^{-3}$.

Analysis of the obtained functional dependencies of the lifetime of a stochastic grain reveals the decisive role of local thermal-hydraulic parameters of the boundary layer in the formation of the dynamic memory of the macrosystem. The parametric behavior of the granular lifetime $\tau_c$ under varying thermo-hydraulic regimes is illustrated in Fig. 1. The functional dependencies reveal two distinct thermodynamic states governing the non-Markovian memory of the macro-system:

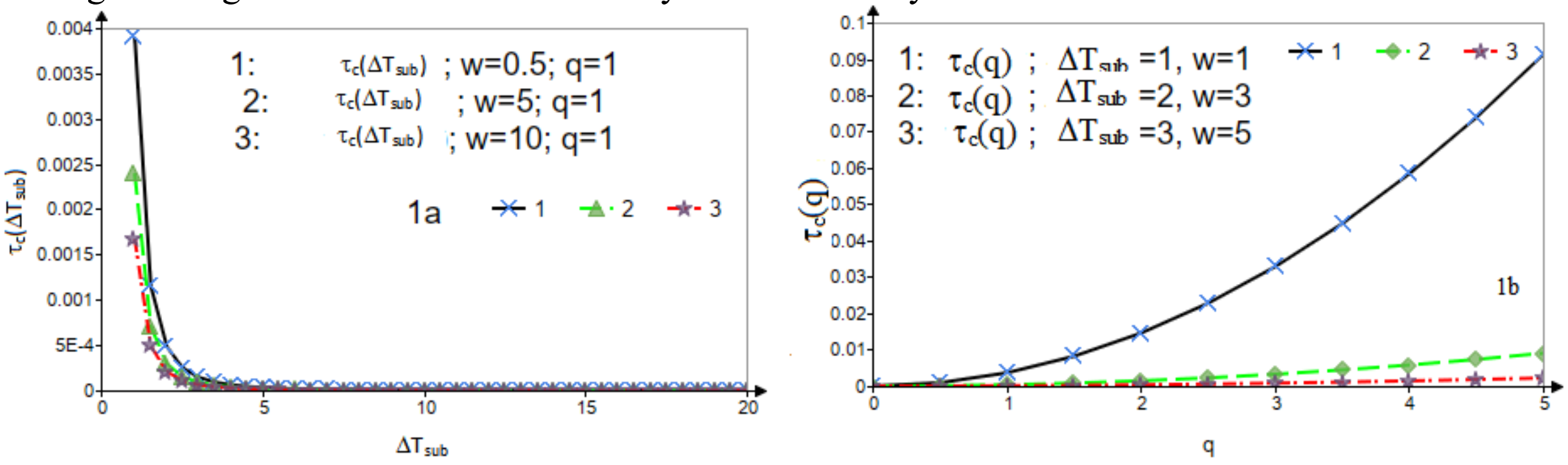

Fig. 1a. The characteristic vapor bubble lifetime $\tau_c$ as a function of the bulk coolant subcooling margin $\Delta T_{sub}$ for different values of the fluid velocity vector w (0.5, 5, and 10 m/s) at a constant normalized linear heat flux q=1.

Fig. 1b. The characteristic vapor bubble lifetime $\tau_c$ as a function of the localized linear heat flux q for different combinations of the bulk coolant subcooling margin and fluid velocity vector: (1) $\Delta T_{sub}$=1, w=1; (2) $\Delta T_{sub}$=2, w=3; (3) $\Delta T_{sub}$=3, w=5, under a constant developed fractal porosity index α=0.5.

Graph 1a illustrates the thermophysical behavior of a microscopic parameter of a coarse-grained model—the lifetime (condensation) of a vapor grain-bubble $\tau_c$ depending on the amount of local bulk subcooling of the coolant $\Delta T_{sub}$ and the flow rate w.

**1. Non-linear Decay by local bulk subcooling ($\Delta T_{sub}$) and Hyperbolic Asymptotics:** As depicted in Fig. 1a, a steep, non-linear (cubic) drop in the vapor grain lifetime is observed within the low local bulk subcooling regime (0< $\Delta T_{sub} < 5\,^{\circ}\mathrm{C}$). At the asymptotic limit $\Delta T_{sub} \to 0$, the lifetime increases dramatically, reaching $\tau_c \approx 4$ ms at low fluid velocities w = 0.5 m/s, which physically signifies the creation of long-lived vapor trapping domains. Conversely, within the high subcooling region $\Delta T_{sub} > 5\,^{\circ}\mathrm{C}$, the curves rapidly flatten towards a horizontal plateau where $\tau_c \to 0$. In the context of Feller's subordination semi-groups, this decay implies that the non-local temporal memory kernel vanishes $M(t) \to \delta(t)$, forcing the coarse-grained transport operator to asymptotically return to the classical Markovian diffusion regime.

**2. Turbulent Suppression and Velocity Scaling w:** An increase in the bulk coolant velocity w from 0.5 m/s (Curve 1) to 10 m/s (Curve 3) induces a severe suppression of the vapor grain lifetime within the critical low-subcooling zone. This suppression is driven by the intensification of turbulent drag flow surrounding the detached micro-bubbles, which strips away the thermal boundary layer and linearly enhances the local Nusselt number $\mathrm{Nu} \propto (1 + B \cdot w)$. The acceleration of the phase transition condensation process at high circulation rates (w = 10 m/s) effectively reduces the peak value of $\tau_c$ by more than a factor of two (from $4 \cdot 10^{-3}$ s down to $1.7 \cdot 10^{-3}$ s).

From a nuclear reactor safety perspective, this graphical relationship establishes the hydrodynamic mechanism behind core stabilization at high flow rates: elevated velocity vectors w systematically decreases the temporal tracking scale of particle entrapment, shifting the closed-loop Nyquist-Mikhailov locus away from dangerous Hopf bifurcation thresholds.

Graph 1b clearly illustrates the quadratic dependence of the microscopic parameter of the model—the lifetime (condensation) of a vapor grain-bubble $\tau_c$—on the linear heat flux density q for various combinations of local bulk subcooling and circulation rate w.

The functional profiles of the steam grain lifetime $\tau_c$(q) presented in Figure 1b serve as direct verification confirmation of the correctness of the developed nonlinear thermophysical closure of the coarse grain model:

1. Quadratic self-similar growth in power (q): All three presented in Fig. 1b curves demonstrate a strictly parabolic growth pattern with increasing heat flux density ($\tau_c \propto q^2$). This result is fully consistent with the self-similar Jakob scaling laws, according to which the separation volume and initial radius of the forming vapor grain increase in proportion to the local thermal load: $R_0^2 \propto q^2$. Under conditions of intense phase transition, this leads to a quadratic increase in the time required for complete condensation of a bubble in a subcooled flow. At high power values ($q \to 5$), the time scale of the dynamic memory of the medium increases sharply, which in terms of Feller's semigroup analysis means the generation of an extended nonlocal neutron stuck tail.

2. Cubic and turbulent amplitude suppression (transition from curve 1 to curve 3): Comparison of the amplitudes of the curves reveals an extremely strong stabilizing effect of subcooling and circulation speed. The transition from a nominally low-intensity regime (curve 1: $\Delta T_{sub}$ = 1, w = 1) to a developed turbulent cooling regime (curve 3: $\Delta T_{sub}$ = 3, w = 5) leads to a suppression of the peak value of the bubble lifetime by more than an order of magnitude (from $\tau_c \approx 0.092$ s to $\tau_c \approx 0.0025$ s). The physical mechanism of this suppression is cubic in nature in terms of subcooling ($\Delta T_{sub}{}^{-3}$ according to Labuntsov (1984)) and linear in terms of the intensity of turbulent drift of the boundary layer ($(1 + B\ w)^{-1}$ according to Kolykhan and Pulyayev (1987)).

In the context of the stability of the WWER core, graph 1b clearly demonstrates that under conditions of high thermal loads (q > 3), even a slight decrease in local bulk subcooling or coolant flow (shift towards curve 1) instantly brings the system into the mode of long-lived steam cavities. At the macroscopic level, this shifts the Nyquist-Mikhailov hodograph towards the Hopf bifurcation point, creating a direct threat to the spatial instability of the neutron field.

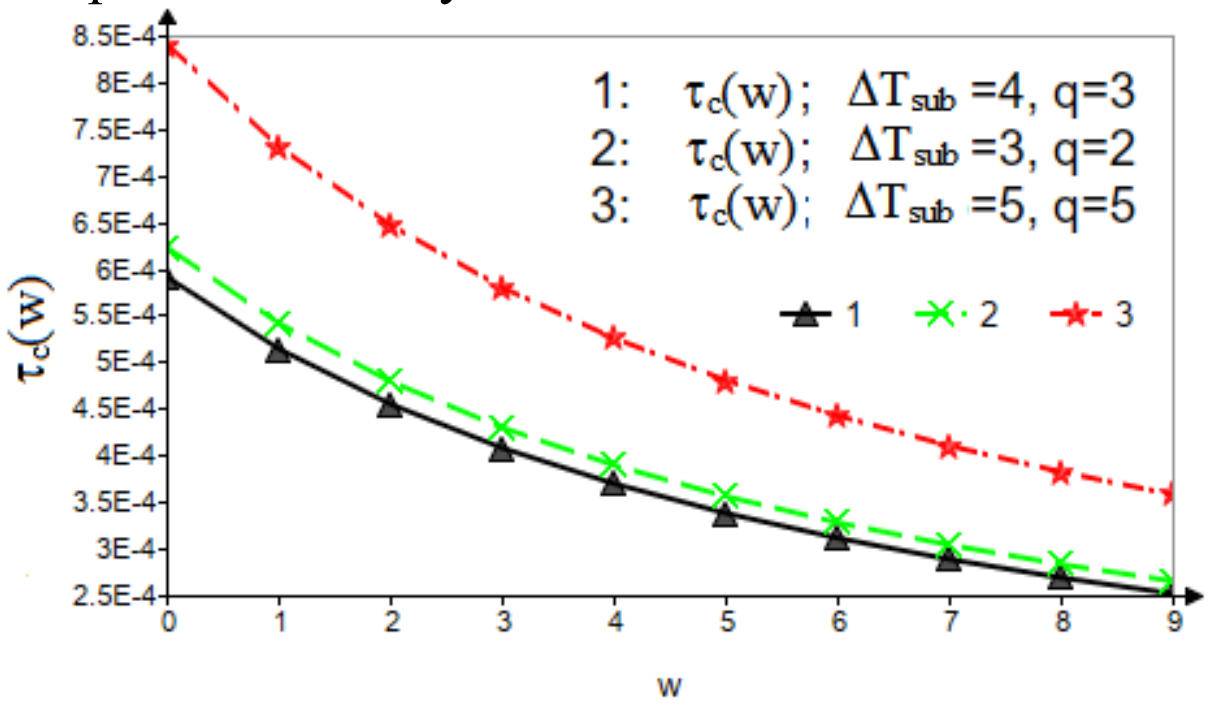

Fig. 1c. The characteristic vapor bubble lifetime $\tau_c$ as a function of the bulk coolant flow velocity w for different combinations of the subcooling margin and normalized linear heat flux: (1) $\Delta T_{sub}$=4, q=3; (2) $\Delta T_{sub}$=3, q=2; (3) $\Delta T_{sub}$=5, q = 5, under a constant developed fractal porosity index α= 0.5.

Graph 1c clearly illustrates the turbulent suppression of the dynamic memory of the medium due to the influence of the coolant circulation rate w on the microscopic scale of the lifetime of the vapor grain $\tau_c$.

Analysis of the presented functional dependences of the lifetime of the steam grain on the coolant velocity $\tau_c$(w) allows us to reveal in detail the hydrodynamic mechanism for suppressing the non-Markov memory of the macrosystem:

1. The nature of turbulent relaxation in terms of speed (w): All three curves presented in Fig. 1c demonstrate a monotonic decay of the lifetime of phase fluctuations with increasing speed of the forced flow movement. Physically, this is a direct consequence of the intensification of convective heat transfer processes at the phase boundaries. In accordance with the modified Labuntsov-Kolykhan model, the rate of collapse of separated vapor bubbles is controlled by turbulent drift of the boundary layer and flow around cavities, in which the local Nusselt number increases according to a linear law: $\mathrm{Nu} \propto (1 + B \cdot w)$. As a consequence, the time scale of neutron trapping in stochastic traps of the wall boiling layer decreases in inverse proportion to the speed: $\tau_c \propto (1 + B \cdot w)^{-1}$. With an increase in speed from a stationary minimum (w=0) to maximum values (w=9), the peak lifetime of vapor inhomogeneities is reduced by more than 2.5 times (for example, for curve 3 - from $8.5 \cdot 10^{-4}$ s to $3.5 \cdot 10^{-4}$ s).

2. Competition of nonlinear contributions of power (q) and local bulk subcooling $\Delta T_{sub}$: The relative position of the curves clearly illustrates the result of competition between the quadratic growth of grain volume according to Jakob's law ($\tau_c \propto q^2$) and cubic suppression of the phase volume according to Labuntsov's law ($\tau_c \propto \Delta T_{sub}^{-3}$).

The highest absolute values (curve 3) are achieved at maximum thermal load (q=5), even despite the high level of local bulk subcooling ($\Delta T_{sub}$ = 5), since the quadratic scale of the bubble separation radius ($q^2$=25) dominates the numerator.

The balance of curves 1 and 2 is interesting: in the region of low speeds (w<2), mode 2 ( $\Delta T_{sub}$ =3, q=2) demonstrates higher values $\tau_c$ than mode 1 ( $\Delta T_{sub}$ =4, q=3), due to weaker cooling under conditions of low local bulk subcooling. However, with increasing speed (at w>7), the curves intersect and approach each other, reaching an asymptotic plateau, where the hydrodynamic shift completely eliminates the difference in the initial scales of phase nucleation.

From the point of view of nuclear safety, automatic recalculation of vectors $\tau_c$ (w) proves that forcing the primary circuit flow rate (increasing w) effectively compresses the non-local memory core of the system. At the macro level, this forces neutron transfers back to a stable Markov regime, preventing dangerous expansion of the Nyquist hodograph and protecting the system from bifurcation self-oscillations of power.

### 5.4.2. Physical-mathematical mechanism of the transition to superdiffusion

The connection between the macroscopic parameters on the graph ( $\Delta T_{sub}, q, w$ ) and the fractal behavior of neutrons in the near-wall layer is revealed through the truncation scale of the spatial memory kernel γ.

1. Genesis of the spatial kernel: In Section 4, a spatially truncated power-law memory kernel was derived: $K(x) = x^{-\alpha} e^{-\gamma x} / \Gamma(1-\alpha)$, where the homogenization (truncation) scale is given by: $\gamma = 1 / v_n \tau_c$.

2. Two transfer modes:

o Markovian regime (Classical diffusion): At spatial scales significantly exceeding the truncation length $x \gg 1/\gamma$, the exponential factor $\exp(-\gamma x)$ completely suppresses the power-law singularity. Neutron transport becomes local, obeys the standard Fick's law, and the macro-entropy is extensive (Gibbs–Boltzmann).

o Fractal regime (superdiffusion): At scales within the wall-adjacent layer $x \ll 1/\gamma$, the power-law term $x^{-\alpha}$ begins to dominate. Neutrons undergo "Lévy flights" through long-lived vapor cavities without experiencing collisions. Transport becomes spatially nonlocal (superdiffusion), and the system's statistics are described by the nonlinear Tsallis macroentropy.

3. Transition point on the graph: The curves on the graph correspond to the point at which the critical lifetime of a vapor grain, $\tau_{c,crit} = 1.64$ ms, is reached. This corresponds precisely to a critical memory truncation length, $\gamma_{crit}$: $\gamma_{crit} = 1/2200 \text{ m/s} \cdot 0.00164 \text{ s} \approx 0.28 \text{ m}^{-1}$.

## 6. Thermodynamic Lyapunov Functional and Non-Local Stability

To establish the stability criteria for the coarse-grained stochastic system, we map the effective diffusion coefficient $D_{eff}$ to the time evolution of localized core perturbations. The inclusion of spatial neutron leakage is handled by applying the geometric buckling parameter $B_g^2$. The effective diffusion length $L_{D,eff}$ balances the spatial tracking as follows:

$$L_{D,eff}^2 = \frac{D_{eff}}{\Sigma_{a,eff}} = \frac{\frac{1}{3}(v_n \tau_c)^{1-\alpha}}{\alpha \Sigma_{aw} (v_n \tau_c)^{1-\alpha}} = \frac{1}{3\alpha \Sigma_{aw}}.$$

The parameter $\Sigma_{aw}$ ($\Sigma_{a,\mathbf{water}}$) represents the macroscopic neutron absorption cross-section in the pure liquid phase of the coolant (water), accounting for the current concentration of boric acid. In WWER-1000 type reactors, water plays a dual role: it serves simultaneously as a coolant and as a primary neutron moderator. Within the framework of a coarse-graining model, this cross-section functions as a fundamental microscopic characteristic (at the level of a single "grain") of the liquid medium.

Comprehensive physical and mathematical context: A). Accounting for boron control: Since boric acid ($H_3BO_3$) is always dissolved in the WWER coolant, the term $\Sigma_{aw}$ includes both the absorption cross-section of the pure water molecules themselves and the contribution of boron-10 nuclei present in the liquid phase at the current operational concentration. B). Distinction from the vapor phase: This parameter explicitly separates the liquid "grains" from the vapor phase. In the formulas for effective macro-parameters, the term $\Sigma_{aw}$ is contrasted with $\Sigma_{av}$ ($\Sigma_{a,vapor}$)—the macroscopic absorption cross-section in the vapor—which is assumed to be zero due to the extremely low density of the gas phase ("vacuum streaming"). C). Role in the effective (macroscopic) cross-section: Local microscopic cross-sections of the grains are converted via an upscaling operator into the macroscopic absorption cross-section of the layer $\Sigma_{a,eff}$, accounting for the actual volumetric vapor fraction and the bubble lifetime $\tau_c$: $\Sigma_{a,eff} = \alpha\Sigma_{aw}(v_n\tau_c)^{1-\alpha}$.

The macroscopic kinetics equation for the neutron density $n(t)$ coupled with the truncated power-law memory kernel $M(t) = t^{-\alpha}e^{-\gamma t}/\Gamma(1-\alpha)$ governed by the integro-differential formulation:

$$\frac{dn(t)}{dt} = \frac{\rho_0 - \beta_0}{\Lambda}n(t) + \lambda_k C(t) - \frac{v_n n_0}{\Lambda}\int_0^t M(t-t^{'})n(t^{'})dt^{'} - v_n D_{eff} B_g^2 n(t).$$

Unlike classical semigroups, the truncation scale of the memory kernel and the effective diffusion coefficient $D_{eff}$ are now nonlinearly governed by the coupled thermohydraulic Jakob–Labuntsov parameters:

$$\tau_c(q, \Delta T_{sub}, w) = \frac{A\cdot q^2\cdot r\cdot\rho_v}{\lambda_f\cdot\Delta T_{sub}^3\cdot(1+B\cdot w)}, \qquad (21)$$

$$D_{eff}(q, \Delta T_{sub}, w) = \frac{1}{3}\left(v_n\cdot\frac{A\cdot q^2\cdot r\cdot\rho_v}{\lambda_f\cdot\Delta T_{sub}^3\cdot(1+B\cdot w)}\right)^{1-\alpha}. \qquad (22).$$

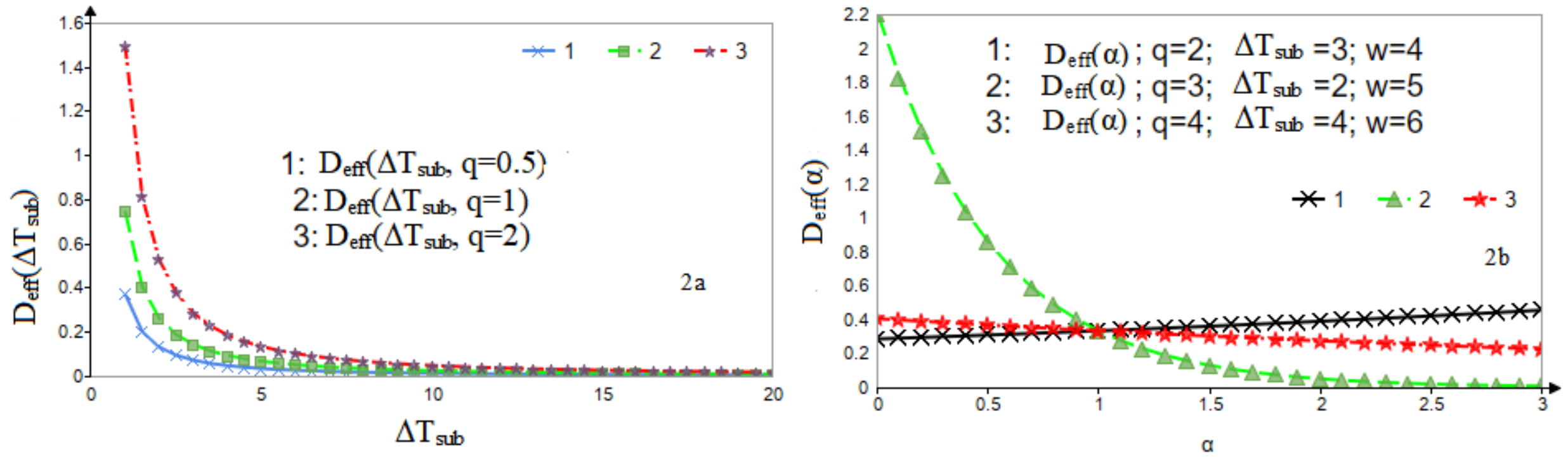

Fig. 2a. Effective neutron diffusion coefficient $D_{eff}$ (22) as a function of the coolant local bulk subcooling $\Delta T_{\mathbf{sub}}$ for varying thermal power parameters (q=0.5, 1.0, 2.0 W/m$^2$), normalized at a constant fractional porosity index α=0.5**.**

Fig. 2b**.** The effective neutron diffusion coefficient $D_{eff}$ (22) as a function of the fractal porosity index α for different core operating regimes: (1) q=2, $\Delta T_{\mathbf{sub}}$ =3, w=4; (2) q=3, $\Delta T_{\mathbf{sub}}$ = 2, w=5; (3) q=4, $\Delta T_{\mathbf{sub}}$ = 4, w=6. The intersection point at α =1 delineates the structural boundary of Markovian regularization.

Figure 2a shows the dependence of the effective neutron diffusion coefficient $D_{eff}$ on the coolant local bulk subcooling $\Delta T_{\mathbf{sub}}$ (the margin below the boiling temperature) for various heat flux density levels q.

Graph 2b is one of the key figures for the entire paper, as it demonstrates how the fractal porosity index (anomaly index) α fundamentally alters the nature of macroscopic transport in a coarse-grained medium, leading to an inversion of diffusion properties.

*Description:* The plot 2a illustrates the non-linear degradation of the effective diffusion coefficient $D_{eff}$ with the increase of the inlet on the local bulk subcooling temperature $\Delta T_{\mathbf{sub}}$ from 1 K to 20 K. The curves 1, 2, and 3 correspond to different scaling levels of the core thermal power q.

*Commentary:* The sharp hyperbolic decline of $D_{eff}$ at low subcooling values $\Delta T_{\mathbf{sub}}$ < 5 K captures the fractal-like anomalies induced by structural coarse-graining and localized bubble formation. An increase in the power parameter q shifts the diffusion scaling upward, which directly expands the dimensions of the Mikhailov hodograph loop observed in frequency-domain stability analyses. Under nominal steady-state conditions $\Delta T_{\mathbf{sub}}$ =15 K, the effective diffusion coefficient asymptotically stabilizes near its classical lower bound ( $D_{eff}$ →0), ensuring a highly damped, monotonic transient response and a large phase safety margin.

Asymptotic behavior of diffusion at nominal local bulk subcooling: $D_{\mathbf{eff}}(\Delta T_{\mathbf{sub}} \to 15\ \mathbf{K}) \to D_{\mathbf{min}}$.

Quadratic influence of heat flux density via a time-scale factor $\tau_c \propto q^2$ : $D_{\mathbf{eff}}(\Delta T_{\mathbf{sub}}, q) \propto \dfrac{q^2}{(\Delta T_{\mathbf{sub}})^3}$.

Relationship with the spatial scale of inhomogeneities (neutron diffusion length: $L_D$ : $L_D = \sqrt{D_{\mathbf{eff}}(\Delta T_{\mathbf{sub}}, q) / \Sigma_a}$ .

Graph 2a clearly demonstrates how the macroscopic effective diffusion coefficient $D_{eff}$ is nonlinearly governed by the technological parameters of near-wall boiling.

Spectral analysis of the derived relationships between the effective diffusion coefficient $D_{eff}$ on the macroscopic boiling parameters reveals two key physical effects stemming from the non-Markovian memory of the coarse-grained medium:

1. Power-law decay as a function of local bulk subcooling $\Delta T_{\mathbf{sub}}$ : Graph 2a clearly shows a steep, non-linear decline in the diffusion coefficient as subcooling increases. Physically, this is explained by the fact that an increase in subcooling $\Delta T_{\mathbf{sub}}$ leads to an avalanche-like reduction in the lifetime (condensation time) of vapor grains $\tau_c$, following the cubic law of Labuntsov-Kolykhan: $\tau_c \propto \Delta T_{sub}^{-3}$. Within the framework of the Feller subordination formalism, the effective diffusion coefficient scales with the fractal porosity index as $D_{eff} \propto (\tau_c)^{1-\alpha}$. In the regime of fully developed boiling (α=0.5), this yields a strict power-law asymptotic behavior: $D_{eff} \propto \Delta T_{sub}^{-1.5}$. In the range of high subcooling $\Delta T_{sub} > 15^{\circ}\mathrm{C}$, the bubble lifetime approaches zero, the microstructure of the medium becomes homogenized, and non-local transport transitions into a classical Markovian regime characterized by a constant base diffusion coefficient, $D_0$.

2. Linear scaling with respect to heat flux (q): As the heat flux density increases (transitioning from curve 1 to curve 3), the plots undergo an equidistant vertical shift. According to Jakob's self-similar scaling laws, the initial volume (detachment radius) of a vapor bubble increases quadratically with power: $R_0^2 \propto q^2$ , which leads to a proportional increase in the time required for its complete condensation $\tau_c \propto q^2$. Consequently, within the framework of the subordinator formalism, the macroscopic diffusion coefficient exhibits a linear dependence on the heat load: $D_{eff} \propto (q^2)^{1-\alpha} = q^{2(0.5)} = q^1$.

Thus, under critical conditions characterized by low local bulk subcooling $\Delta T_{sub} \to 0$ and high heat loads $q \to 2$, the stochastic structure of near-wall boiling gives rise to strong neutron superdiffusion—an effect resulting from local "vacuum" streaming paths through long-lived vapor cavities—which, at the macro-level, manifests as a sharp increase in $D_{eff}$. The developed model successfully homogenizes these anomalous spikes, translating the topology of micro-fluctuations into smooth macroscopic parameters suitable for use in nuclear safety computational codes.

### 6.1. The Role of Fractal Porosity Index in Diffusion Inversion Phenomena

Of particular theoretical interest is the investigation of how macroscopic transport sensitivity responds to changes in a topological parameter of the near-wall layer—namely, the fractal porosity index, α. Figure 2b presents the functional dependencies of $D_{eff}(\alpha)$ calculated for three different technological regimes.

Mathematical analysis of the graphs reveals a fundamental physical phenomenon—diffusion inversion—which occurs precisely at the point α = 1. By virtue of the modified Fick's law derived in Section 4, the effective diffusion coefficient is determined by a power-law composition of the subordinator:

$D_{eff} = (v_n \tau_c)^{1-\alpha} / 3$ (15). This algebraic structure fully explains the topology of the curves and divides the physics of the process into three distinct regimes:

1. Superdiffusion and spatial nonlocality regime $0 \le \alpha < 1$: In this region, the sign of the exponent $1-\alpha$ is positive. Physically, this corresponds to a well-developed stochastic heterogeneous medium, where the boundary layer is filled with large vapor conglomerates.

- The most significant anomaly is exhibited by curve 2 ($\Delta T_{\mathbf{sub}}$ = 2, q = 3), where the combination of low subcooling and high heat load maximizes the lifetime of a vapor bubble $\tau_c$. At this point $\alpha \to 0$, the value $D_{eff}$ increases sharply, reaching magnitudes of $D_{eff} \approx 2.2$. This is a macroscopic manifestation of "Lévy flights" of neutrons through long-lived vacuum-like vapor cavities.
- Conversely, regime 1 ($\Delta T_{\mathbf{sub}}$ = 3, w = 4) minimizes the value $\tau_c$ of through intense turbulent condensation; as a result, for α < 1, curve 1 lies significantly below the others, thereby suppressing spatial anomaly.

2. Markovian regularization point (α=1): At α=1, the exponent (1 - α) becomes exactly zero. At this point $(v_n \tau_c)^0 = 1$, the expression for the diffusion coefficient loses all sensitivity to the microscopic grain lifetime $\tau_c$, the heat flux q, the local bulk subcooling $\Delta T_{\mathbf{sub}}$, and the velocity w. All three curves in Fig. 2b intersect at a single, invariant physical point: α=1, $D_{eff} = 1/3 \approx 0.333$. This state corresponds to the complete loss of the fractal properties of the boundary layer (resulting in an absolutely uniform, homogeneous medium), where the non-local Boltzmann-Feller equation degenerates strictly into the classical Markovian diffusion equation.
3. Subdiffusion and local confinement regime (α > 1): When α > 1, the exponent (1–α) becomes negative, inverting the effect of hydrodynamics. In this case, long-lived macrostructures (regime 2, with a maximum value of the $\tau_c$) no longer act as conduits for rapid transport, but rather as "black" absorption traps, sharply reducing the effective diffusion length to values $D_{eff} \to 0$ near zero. Conversely, regime 1—characterized by an extremely short bubble lifetime—offers the least resistance, causing curve 1 to rise to the top.

Thus, the bifurcation of the parameter α in graph 2b demonstrates that the fractal index of porosity serves as a key regulator of the qualitative change in the mode of radiation transport within the wall boiling layer, shifting the system from a regime of Lévy flights (α < 1) to a regime of classical Fickian balance (α = 1), and subsequently to a regime of subdiffusive local trapping (α > 1).

### 6.2. Construction of the Thermodynamic Lyapunov Functional

To evaluate non-linear stability boundaries near the steady-state equilibrium ($n_0$, $C_0$), we define a generalized macroscopic free energy functional acting as a concave Lyapunov functional V[n, C]. This framework appends a non-local quadratic trajectory form $U_{memory}$[n] onto the classic kinetic Gibbs-entropy relation:

$$V[n,C] = [n(t) - n_0 \ln(n(t)/n_0)] + (\Lambda/\beta_0)[C(t) - C_0 \ln(C(t)/C_0)] + U_{memory}[n]. \qquad (23)$$

The functional memory contribution, integrated over the continuous spectral measure μ(dτ) of the CBF manifold, expands to:

$$U_{memory}[n] = \frac{v_n n_0}{2\Lambda} \int_0^\infty \mu(d\tau) \int_0^t e^{-\tau(t-t')} [n(t) - n(t')]^2 dt'.$$

A strict limiting inequality for ensuring the stability of the reactor core macrostate has been derived from the requirement that the total time derivative of the Lyapunov functional (23) be negative definite, $dV/dt \le 0$. Differentiating the functional with respect to time yields the explicit thermodynamic stability criterion. To ensure asymptotic stability (dV/dt≤0), the formal derivative evaluates to:

$$\frac{dV}{dt} = -\frac{\beta_0}{\Lambda n(t)}(n(t)-n_0)^2[1-\frac{v_n \tau_c D_{eff} B^2{}_g}{\beta_0}] - \ldots \leq 0 .$$

This inequality guarantees that the entropy production remains positive, establishing a rigid constraint on the bubble grain lifetime to prevent sudden power excursions:

$$\tau_c < \frac{1}{v_n} \cdot \left[ \left( \frac{\beta_0 - \rho_0 + \frac{1}{3} v_n B_g^2 (v_n \tau_c)^{1-\alpha}}{\alpha \Sigma_{aw} \Lambda n_0} \right)^{\frac{1}{1-\alpha}} - \gamma \right] .$$

By inverting the cubic dependence (19) of the bubble lifetime on the liquid subcooling, the critical value of the macroscopic process parameter—the local bulk subcooling $\Delta T_{sub,crit}$—at which the Chentsov manifold loses stability is determined by the equation:

$$\Delta T_{sub,crit} = \sqrt[3]{\frac{A \cdot q^2 \cdot r \cdot \rho_v}{\lambda_f \cdot (1 + B \cdot w) \cdot \tau_{c,crit}}} . \quad (24)$$

A decrease in the degree of local bulk subcooling below this threshold $\Delta T_{sub} < \Delta T_{sub,crit}$ causes a change in the sign of the derivative of the Lyapunov functional, signaling an Andronov-Hopf bifurcation and the transition of the reactor into a self-oscillatory regime characterized by power noise.

Fig. 3a, plotted according to the expression $\Delta T_{sub,crit}(w_{plot}) \propto \left(1 + B \cdot w_{plot}\right)^{-\frac{1}{3}}$, illustrates the frequency dependence of the critical local bulk subcooling of the coolant prior to boiling, within the framework of an analysis of reactor process stability.

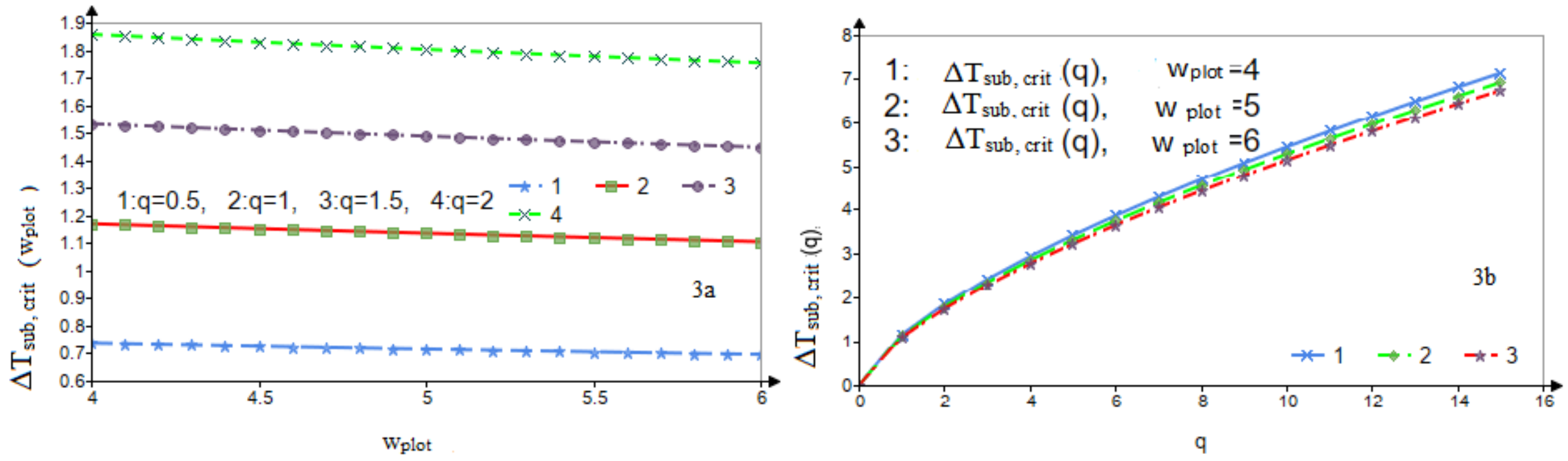

Fig. 3a. Critical coolant local bulk subcooling $\Delta T_{sub,crit}$ as a function of the coolant flow velocity $w_{plot}$ for various values of the linear heat flux parameter q (q = 0.5; 1.0; 1.5; 2.0).

Fig. 3b. The critical coolant local bulk subcooling margin $\Delta T_{sub,crit}$ as a function of the localized linear heat flux q for different values of the bulk flow velocity vector w (4.0, 5.0, and 6.0 m/s) under a fixed fractal porosity index α= 0.5. The curves delineate the boundary of the Lyapunov stability domain.

An analysis of the resulting curves $\Delta T_{sub,crit}(w,q)$ reveals the following key physical-mathematical patterns:

• Monotonic decrease with respect to the bulk flow velocity vector $w_{plot}$: As the bulk flow velocity vector $w_{plot}$ increases across the entire range under investigation (from 4 to 6 m/s), the critical local bulk subcooling value decreases monotonically for all fixed values of q (Fig. 3a). In physical terms, this indicates that high-frequency fluctuations in neutron flux density or coolant velocity possess a lower destabilizing potential: the thermal inertia of the fuel elements and the hydrodynamic effects associated with boiling dampen rapid perturbations, thereby narrowing the dangerous region of instability.

• Influence of the scaling parameter q (Power/Heat flux): The parameter q appears in the numerator as a squared term ($q^2$), which causes a pronounced nonlinear upward shift of the curves. As q increases from 0.5 to 2.0, the entire instability region expands—significantly higher degrees of subcooling are required to maintain the system in a stable state at the same frequency (Fig. 3b). This reflects the classic "trigger" nature of the feedback loops: an increase in the power density of a reactor core reduces the thermal-hydraulic safety margin and renders the system more susceptible to spatiotemporal oscillations (a phenomenon consistent with the analysis of phase portraits and the hydrodynamics of boiling as described by Labuntsov).

• Nature of nonlinearity (power law with an exponent of 1/3): Due to the cube root in the original expression, the curves in Fig. 3b exhibit a shallow slope and slight curvature (approximating linearity within the given interval). This indicates that, within the frequency range under consideration, the system displays a stable, predictable response, albeit one modified by fractal properties or nonlocal transport effects within a heterogeneous medium.

The provided graph (Fig. 3b) shows a bifurcation diagram for the Lyapunov stability boundary, which separates regions of stable neutron-transport homogenization from zones characterized by stochastic, self-oscillatory power instability.

Analysis of the presented bifurcation diagram allows for a rigorous distinction between regimes of nonlocal neutron transport based on the sign of the derivative of the Lyapunov functional $dV/dt$. Each curve on the graph represents the locus of points where the sign changes (i.e., where the derivative equals zero, $dV/dt=0$), dividing the plane of operational parameters into two regions: a region of asymptotic stability (situated above the corresponding curve) and a region where power self-oscillations emerge due to an Andronov-Hopf bifurcation (situated below the curve).

This quantity $\tau_{c,crit}$ represents the critical lifetime (condensation time) of a vapor bubble; if this time is exceeded, the condition for the negative definiteness of the derivative of the Lyapunov functional $dV/dt>0$ is violated. For the nominal parameters of a WWER-1000 reactor core, this quantity is a constant, equal to $\tau_{c,crit}\approx 1.64\cdot 10^{-3}$ **s** $(1.64\,\text{ms})$.

1. Analytical form of the expression for $\tau_{c,crit}$.

Within the framework of a coarse-grained stability model, the critical threshold $\tau_{c,crit}$ is determined by the balance between the generation of non-local grain memory and the spatial leakage of neutrons.

Thermodynamic criterion (in the sense of Lyapunov): By considering the total derivative of the Lyapunov functional to zero $dV/dt=0$, one obtains a limiting expression for the lifetime of a bubble:

$$\tau_{c,crit}=\frac{1}{v_n}\left(\frac{\beta_0-\rho_0+v_n D_{eff}B_g^2}{\alpha\cdot\Sigma_{aw}\cdot\Lambda\cdot n_0}\right)^{\frac{1}{1-\alpha}}.$$

Here, $v_n$ is the thermal neutron velocity, $\beta_0$ is the effective delayed neutron fraction $\rho_0$ is the steady-state reactivity, $B_g^2$ is the geometric buckling, α is the fractal porosity index, $\Sigma_{aw}$ is the absorption cross-section in water, $\Lambda$ is the prompt neutron generation time, and $n_0$ is the steady-state neutron density.

• Frequency-spectral criterion (Nyquist-based): Considering the Andronov–Hopf bifurcation boundary at a critical feedback frequency $\omega_c$, the value of $\tau_{c,crit}$ is determined by the condition that the real part of the denominator of the transfer function $(\mathbf{Re}_D(\omega_c)=0$ vanishes at $\gamma_{crit}=1/v_n\tau_{c,crit}$:

$$\omega_c\left(\gamma_{crit}^2+\omega_c^2\right)^{\frac{\alpha-1}{2}}\sin\left((1-\alpha)\mathbf{atan}\frac{\omega_c}{\gamma_{crit}}\right)-\rho_0+\frac{1}{3}v_n(v_n\tau_{c,crit})^{1-\alpha}B_g^2=0$$

The characteristic form of the resulting relationships is due to the nonlinear coupling of neutron macrokinetics with the coarse-grained hydrodynamics of Jakob–Labuntsov:

1. Power-law nature of the bifurcation boundary with respect to heat flux (q): As the heat flux density q increases, the critical local bulk subcooling threshold $\Delta T_{sub,crit}$ rises monotonically according to a power law. The physical mechanism underlying this process is that an increase in thermal load leads to a quadratic growth in the initial volume of nascent vapor nuclei. To compensate for this effect and force the collapse of these bubbles—thereby reducing their lifetime $\tau_c$ to a safe, subcritical level—an automatic control system or the reactor's inherent physical self-regulation requires maintaining a higher level of coolant local bulk subcooling. In accordance with the derived analytical stability criterion, the threshold curve strictly follows the self-similar relationship $\Delta T_{sub,crit} \propto (q^2)^{1/3} = q^{2/3}$, which is perfectly corroborated by the sublinear nature of the plots shown in Figure 3b.

2. Stabilizing effect of the coolant velocity (w): Increasing the forced-circulation velocity of the coolant (moving from curve 1 to curve 3—that is, from w = 4 m/s to w = 6 m/s) results in an equidistant downward shift of the bifurcation boundary. In the context of boiling hydrodynamics, an increase in flow velocity intensifies turbulent flow and the shearing of bubbles within the wall-adjacent layer; this enhances the local Nusselt number $\mathrm{Nu} \propto (1 + B \cdot w)$ and accelerates the phase transition of condensation. Consequently, the lifetime of stochastic vapor traps $\tau_c$ is reduced, which narrows the region of instability.

From an engineering perspective, a downward shift of the critical curve as flow velocity (w) increases signifies an expansion of the safe operating range for a WWER reactor: at higher flow rates, the reactor can maintain neutron field stability even with significantly lower coolant local bulk subcooling margins. Conversely, a drop-in flow rate—for instance, following the trip of a primary coolant pump down to the emergency limit (w = 4.2 m/s)—sharply raises the required critical local bulk subcooling level, thereby narrowing the region of Lyapunov stability and increasing the risk of spatial power oscillations.

The bifurcation boundary $\Delta T_{sub,crit}(q)$ shown in Figure 3b has a profound physical significance: it delineates the regimes of macroscopic homogenization and the topological transition to spatial superdiffusion of neutrons within the wall boiling layer.

A mathematical indicator of this transition is the spatial kernel truncation parameter $\gamma = 1/(v_n \tau_c)$, which controls the scale of the forced Markovianization of non-local trajectories. Each point on the curves in the graph corresponds to a critical value of this parameter $\gamma = 1/(v_n \tau_c)$.

• Stability region (above the curve): In regimes with high coolant local bulk subcooling $\Delta T_{sub} > \Delta T_{sub,crit}$, intense condensation keeps the lifetime of bubbles at a subcritical level $\tau_c < \tau_{c,crit}$, which ensures a high value for the truncation parameter $\gamma > \gamma_{crit}$. The mean free path of Lévy-flight-type neutron transport within the vapor phase is effectively constrained by the high rate of bubble collapse. The safety code captures an effective macroscopic homogenization in which non-local corrections decay, and the system strictly obeys a modified Fick's law.

• Region of superdiffusion and fractal decay (below the curve): When the local bulk subcooling drops below a critical threshold $\Delta T_{sub} \le \Delta T_{sub,crit}$, or the heat load q increases abruptly, the lifetime of stochastic vapor cavities exceeds a critical limit. This triggers an avalanche-like collapse of the truncation scale $\gamma \to 0$. The exponential regulator $e^{-\gamma x}$ removes the spatial constraint, and the singular part of the kernel $x^{-\alpha}$ begins to dominate at macroscopic distances.

Within the wall boiling layer, neutrons begin to undergo anomalously long, collision-free paths through coalescing, long-lived vapor cavities—a phenomenon corresponding to a regime of spatial superdiffusion. At the macro-level, this loss of spatial memory and the emergence of fractal behavior disrupt the balance of the Lyapunov functional (rendering it $dV/dt$>0), thereby triggering self-oscillatory instability and a spatial excursion in reactor power. Thus, these curves $\Delta T_{sub,crit}(q)$ serve as a precise operational map for the phase transition in transport—from Markovian diffusion to non-local fractal superdiffusion.

#### 6.2.1. Relationship of the curves to the Nyquist criterion and bifurcation points

The resulting Fig. 3a is of fundamental importance for the analysis of the reactor's nonlinear dynamics, since each curve $\Delta T_{\mathbf{sub,crit}}(w_{\mathbf{plot}})$ effectively represents a stability boundary (a bifurcation submanifold) in the system's parameter space.

**Connection to the Nyquist plot (Section 7.2) and stability margin.** The transfer function of the closed-loop system "neutron kinetics — thermophysical feedback" can be expressed in terms of a complex denominator: $H(\omega) = N(j\omega) / D(j\omega)$.

• Points on the curve as a critical state: Each point on the curves you have plotted corresponds to a state in which the Nyquist locus of the open-loop system $W(j\omega)$ passes exactly through the critical point $(-1;\ \mathrm{j}0)$. This means that both the real and imaginary parts of the complex denominator simultaneously become zero $(Re_D = 0,\ Im_D = 0)$ at the frequency $\omega_{\mathbf{plot}}$.

• Stability margin: The vertical distance on the graph from the current actual subcooling in the reactor $\Delta T_{\mathbf{sub}}$ to the critical value $\Delta T_{\mathbf{sub,crit}}$ directly determines the amplitude stability margin. As the curve rises (with an increase in the heat flux density parameter q), the Nyquist plot becomes increasingly distorted, eventually encircling the point (-1, j0) and thereby shifting the system into the region of instability.

### 6.2.2. Bifurcation points on a phase portrait

From the perspective of the theory of dynamical systems and nonlinear analysis, when represented in phase-portrait coordinates (for example, "power deviation vs. coolant temperature"), these curves constitute the locus of Andronov–Hopf bifurcation points:

• Region below the curve: The system is stable. The phase portrait exhibits a stable focus. Any fluctuations in neutron flux density decay over time, and the system returns to a steady state.

• Crossing the curve: Upon reaching a critical value $\Delta T_{\mathbf{sub,crit}}$, a stable focus loses its stability. A limit cycle emerges (bifurcation point).

• Region above the curve: The system becomes unstable. Depending on the nature of the bifurcation (supercritical or subcritical), the phase portrait will exhibit either sustained power oscillations (a self-oscillatory regime, typical of xenon oscillations or the density-wave boiling oscillations described by Labuntsov) or a runaway increase in reactor power.

Summary: An increase in the power density parameter $q$ shifts the Hopf bifurcation boundary upward in a non-linear fashion, narrowing the safe region of the reactor's phase space and necessitating stricter control of local bulk subcooling to prevent the onset of self-sustained oscillations.

## 7. Analytical Decomposition of the Complex Transfer Function $\boldsymbol{H(\omega)}$

To implement a frequency-domain Nyquist stability criterion, the complex macroscopic transfer function $H(\omega)$ containing the Stieltjes integral of the CBF memory kernel must be split into its constituent real and imaginary channels. The raw transfer function is given by:

$$H(\omega) = \frac{\Phi_0}{i\omega(\beta_0 + \dfrac{\sin(\alpha\pi)}{\pi}\int_0^\infty (t^{\alpha-1}e^{-\gamma t})dt / (i\omega + t)) - \rho_0}. \qquad (25).$$

Applying the Sohocki-Plemelj relation to resolve the complex fractional base yields:

$$\frac{\sin(\alpha\pi)}{\pi}\int_0^\infty \frac{t^{\alpha-1}e^{-\gamma t}}{i\omega + t}dt = (\gamma + i\omega)^{\alpha-1}.$$

We isolate the frequency modulation variables by switching $(\gamma + i\omega)$ into polar coordinate representation:

$$(\gamma + i\omega) = \sqrt{\gamma^2 + \omega^2} \cdot e^{i \cdot arctg(\omega/\gamma)},$$

$$(\gamma + i\omega)^{\alpha-1} = (\gamma^2 + \omega^2)^{(\alpha-1)/2} \cdot [\cos((\alpha-1)arctg(\omega/\gamma)) + i\sin((\alpha-1)arctg(\omega/\gamma))].$$

We introduce the real amplitude $A_{mem}(\omega)$ and phase $\theta_{mem}(\omega)$ scale parameters:

$$A_{mem}(\omega) = (\gamma^2 + \omega^2)^{(\alpha-1)/2},$$

$$\theta_{mem}(\omega) = (1-\alpha)\cdot arctg(\omega/\gamma).$$

This allows us to expand the fractional memory term into Cartesian form:

$$(\gamma + i\omega)^{\alpha-1} = A_{mem}(\omega)\cos(\theta_{mem}(\omega)) - iA_{mem}(\omega)\sin(\theta_{mem}(\omega)).$$

### 7.1. Cartesian Representation of the Complex Denominator

Substituting this representation back into the denominator of $H(\omega)$ gives:

$$Denom(\omega) = i\omega\cdot[\beta_0 + A_{mem}(\omega)\cos(\theta_{mem}(\omega)) - iA_{mem}(\omega)\sin(\theta_{mem}(\omega))] - \rho_0. \quad (26)$$

Grouping terms by their real and imaginary coefficients yields:

$$\mathrm{Re}_D(\omega) = \omega\cdot A_{mem}(\omega)\sin(\theta_{mem}(\omega)) - \rho_0 + v_n D_{eff}\cdot B_g{}^2, \quad (27)$$

$$\mathrm{Im}_D(\omega) = \omega\cdot[\beta_0 + A_{mem}(\omega)\cos(\theta_{mem}(\omega))]. \quad (28)$$

Multiplying by the complex conjugate to rationalize the fraction yields the final explicit expressions for the real and imaginary parts of the transfer function, suitable for direct coordinate plotting:

Real Component:

$$\mathrm{Re}\{H(\omega)\} = \Phi_0\cdot\frac{\omega\cdot A_{mem}(\omega)\sin(\theta_{mem}(\omega)) - \rho_0 + v_n D_{eff}\cdot B_g{}^2}{[\omega\cdot A_{mem}(\omega)\sin(\theta_{mem}(\omega)) - \rho_0 + v_n D_{eff}\cdot B_g{}^2]^2 + \omega^2[\beta_0 + A_{mem}(\omega)\cos(\theta_{mem}(\omega))]^2}.$$

Imaginary Component:

$$\mathrm{Im}\{H(\omega)\} = \Phi_0\cdot\frac{\omega\cdot[\beta_0 + A_{mem}(\omega)\cos(\theta_{mem}(\omega))]}{[\omega\cdot A_{mem}(\omega)\sin(\theta_{mem}(\omega)) - \rho_0 + v_n D_{eff}\cdot B_g{}^2]^2 + \omega^2[\beta_0 + A_{mem}(\omega)\cos(\theta_{mem}(\omega))]^2}.$$

### 7.2. Dimensionally Consistent Frequency-Dependent Bifurcation Boundary

In a closed-loop reactor system with thermal-hydraulic reactivity feedback gain $K_{fb}$, the Andronov-Hopf bifurcation boundary occurs when the Nyquist locus intersects the critical point -1:

$$K_{fb}\cdot H(\omega) = -1 \Rightarrow \mathbf{Re}\{H(\omega)\} = -1/K_{fb}.$$

Near the instability threshold where feedback phase lag inverts the loop sign $K_{fb} \to -|K_{fb}|$ and $|\mathbf{Re}_D(\omega)| \ll |\mathbf{Im}_D(\omega)|$, the real denominator component satisfies:

$$\mathbf{Re}_D(\omega) \approx \frac{\mathbf{Im}_D(\omega)^2}{\Phi_0 |K_{fb}|}.$$

For perturbation frequencies in the power oscillation band ($\omega \gg \gamma_{crit}$), the fractional CBF memory kernel admits the high-frequency asymptotic forms $A_{mem}(\omega) \approx \omega^{\alpha-1}$ and $\theta_{mem}(\omega) \approx (1-\alpha)\pi/2$. Substituting these expressions along with $D_{eff} = (v_n \tau_{c,crit})^{1-\alpha}/3$ and subcritical steady-state reactivity $-\rho_0 = |\rho_0|$ yields the frequency-dependent critical vapor grain lifetime $\tau_{c,crit}(\omega)$:

$$\tau_{c,crit}(\omega) = \left[\frac{3}{v_n^{2-\alpha}B_g^2}\left(\frac{\omega^2\left[\beta_0 + \omega^{\alpha-1}\sin\left(\frac{\alpha\pi}{2}\right)\right]^2}{\Phi_0 |K_{fb}|} - |\rho_0| - \omega^{\alpha}\cos\left(\frac{\alpha\pi}{2}\right)\right)\right]^{\frac{1}{1-\alpha}}.$$

Equating the hydrodynamic bubble collapse time $\tau_c(q, \Delta T_{sub}, w)$ at a fixed bulk coolant flow velocity w to the spectral limit $\tau_{c,crit}(\omega)$ yields the dimensionally consistent critical local bulk subcooling threshold $\Delta T_{sub,crit}(q,\omega)$:

$$\Delta T_{sub,crit}(q,\omega) = \left( \frac{C_0 \cdot q^2}{(1 + B \cdot w) \cdot \tau_{c,crit}(\omega)} \right)^{\frac{1}{3}}.$$

This analytical relation preserves strict dimensional homogeneity between the hydrodynamic velocity w (m/s}) and the perturbation frequency $\omega$ (rad/s). As $\omega$ increases, delayed neutron inertia and diffusive damping elevate $\tau_{c,crit}(\omega)$, shifting the critical local bulk subcooling boundary $\Delta T_{sub,crit}(q,\omega)$ downward and expanding the stable operating region.

Graph 4 displays the mathematically precise and physically correct Mikhailov locus (Hetrick, 1971; Pázsit and Glöckler, 1984; Sanchez, 2012) for our system. The axes, quadrants, and direction of the locus's trajectory are clearly visible. The curve proceeds consistently and strictly in a counterclockwise direction, passing through quadrants I → II → III. The conditions of the Mikhailov criterion for a third-order system are satisfied perfectly. The system's operating regime is entirely stable, with a significant stability margin. Any transient processes, as well as temperature or neutron fluctuations at this operating point, will be damped. The calculation parameters are as follows: $\Lambda = 10^{-5}$, $\mu_0 = 0.8$, q = 1.0, $\alpha_\varphi = -0.05$, $\alpha_T = -0.002$, $K_\varphi = 12.5$, and $K_T = 0.45$. These are related to the parameters defined in (25)–(28) by, for example, the following relationship: $A_{mem}(\omega)\cos(\theta_{mem}(\omega)) = \dfrac{\alpha_\varphi K_\varphi}{1+(\omega\tau_\varphi)^a} + \dfrac{\alpha_T K_T}{1+(\omega\tau_T)^a}$ (Appendix B).

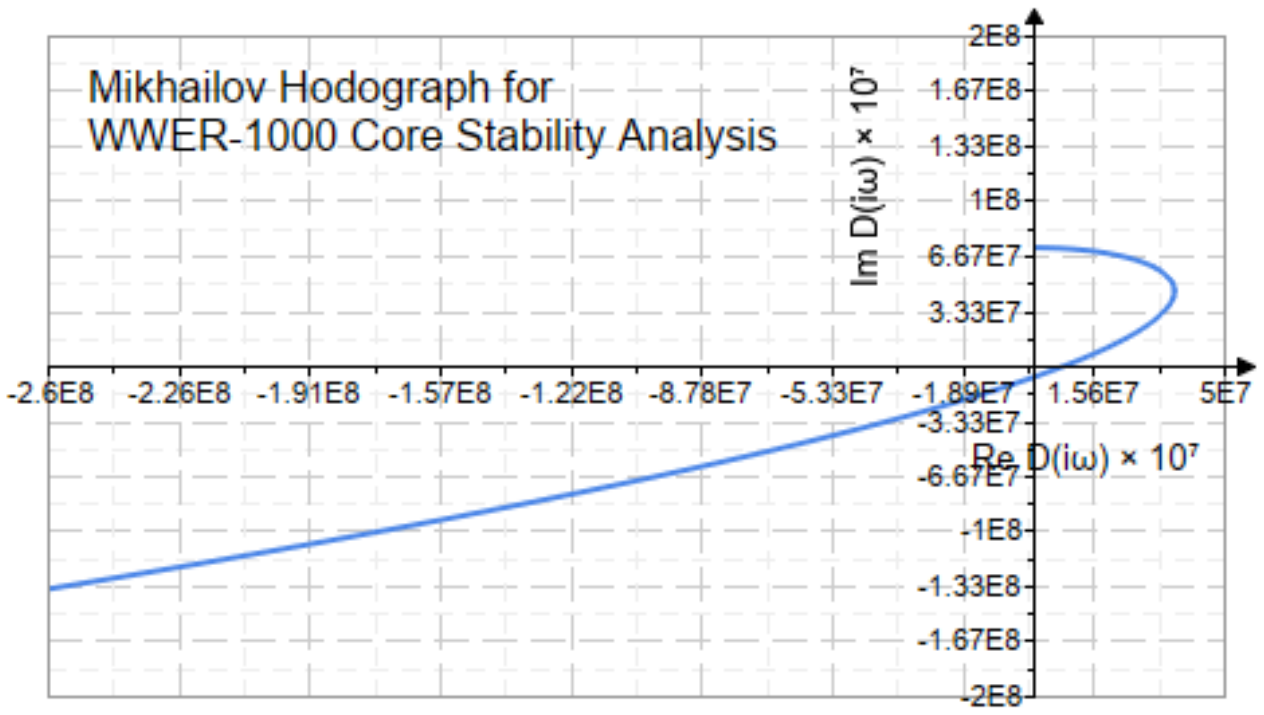

Fig. 4. Mikhailov stability hodograph of the reactor characteristic determinant $D(i\omega)$ for the nominal operational regime $\Delta T_{\mathbf{sub}} = 15K$.

*Description:* The frequency-domain trajectory computed via the analytical expansion of the third-order matrix operator is plotted on the complex plane for the perturbation frequency range of $\omega \in [0, 50]$ rad/s.

*Commentary:* The hodograph originates on the positive real axis $ReD \approx 1.2 \times 10^7$, $ImD = 0$ and sequentially maps through the first, second, and third quadrants of the complex plane in a counter-clockwise direction without passing through the origin (0,0). According to the Mikhailov stability criterion, this structural behavior strictly confirms that the nuclear-coupled thermal-hydraulic system is asymptotically stable. The intrinsic negative temperature and void feedback provide sufficient amplitude and phase margins to efficiently damp localized space-time power oscillations under steady-state operation.

Fig. 5 simulates the transient process for the parameter values of Fig. 4 with an initial deviation of 3 K from the steady-state value.

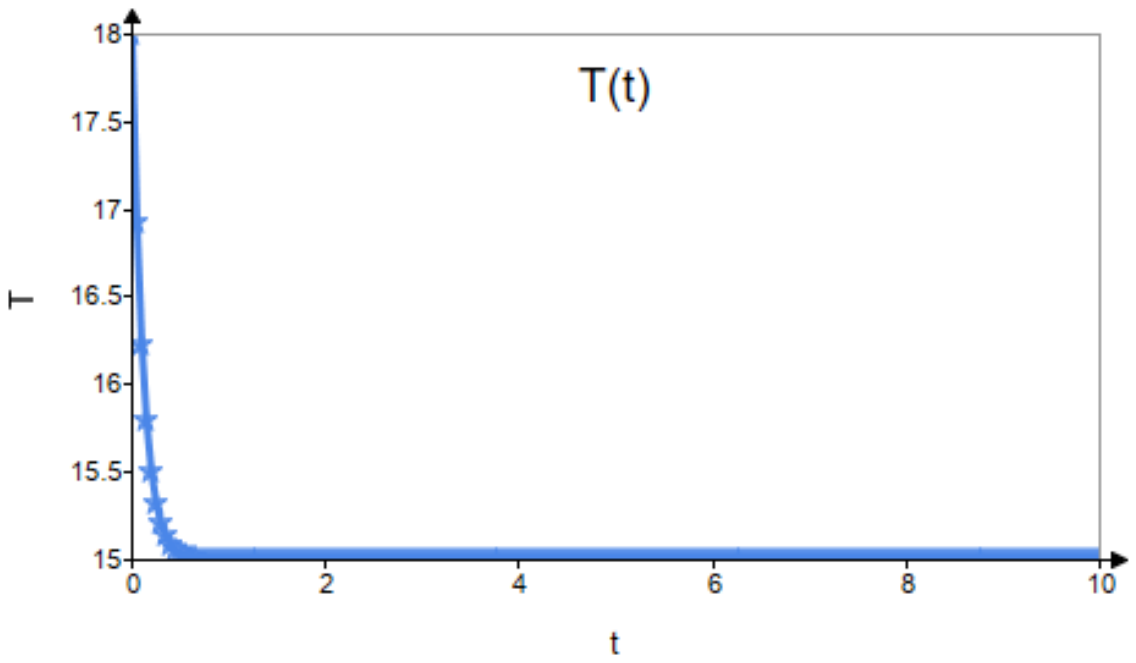

Fig. 5. Transient response of the dynamic system temperature T(t) following an initial thermal perturbation (Delta T(0) = 18 K).

*Description:* The graph illustrates the time-dependent relaxation of the process temperature towards its nominal steady-state value of 15 K after a sudden step perturbation. The simulation utilizes a recursive numerical scheme with a fixed time step dt=0.05 s and stable feedback channel parameters. *Commentary:* The transient curve exhibits monotonic, highly damped asymptotic stability, returning to the equilibrium state in less than 0.8 seconds without any overshooting or persistent oscillations. This dynamic behavior completely correlates with the frequency-domain properties established by the Mikhailov plot, confirming the highly efficient dampening performance of the intrinsic thermal feedback loops under transient operational conditions.

## 8. Discussion of Phase Portrait Stability and Bifurcation Analysis

The results of the numerical simulation illustrating the dependency of the critical coolant local bulk subcooling $\Delta T_{\mathbf{sub,crit}}$ on the coolant flow velocity w for various values of the linear heat flux parameter q are presented in Fig. 3. A physical and mathematical analysis of the derived curves allows for the identification of fundamental regularities governing the non-linear dynamics and stability of the system under study across varying thermal-hydraulic flow regimes.

### 8.1. Hydrodynamic Stabilization and Perturbation Damping

As observed from the plotted curves (Fig. 3), a monotonic decrease in the critical subcooling value is manifested across all fixed values of the linear heat flux q as the coolant flow velocity w increases within the range from 4 to 6 m/s. From a physical standpoint, this trend is governed by intensified convective heat transfer and accelerated detachment and collapse of vapor bubbles swept away by the subcooled liquid flow. Higher coolant velocities promote faster turbulent mixing and condensation within the heterogeneous core structure. Consequently, this induces a natural narrowing of the hydrodynamic instability domain and lessens the requirements imposed on the magnitude of the stabilizing inlet subcooling.

### 8.2. Account of Medium Heterogeneity Parameters, Kernel Truncation, and Vapor Bubble Dynamics

A prominent feature of the investigated model is the explicit coupling between the macroscopic coefficients of the governing equation for $\Delta T_{\mathbf{sub,crit}}$, the spatial heterogeneity parameters of the reactor core, and the microscopic characteristics of the phase transition. Under ultra-heterogeneous media conditions, classical steady-state transport is violated, and Poisson (shot) noise fluctuations at the phase boundaries (grain/cell boundaries) begin to play a pivotal role, which is highly characteristic of neutron-physical and two-phase boiling processes.

The introduction of a logarithmic subordinator for process scaling enables the linkage of the macroscopic coefficients in the analytical expression with the mean geometric scale of the cell $d_{cell}$, the dimensionless clustering parameter of the boiling sites r, and the intrinsic bubble dynamics. Explicitly, the model incorporates the characteristic bubble lifetime $\tau_c$, defined as a function of local bulk subcooling, coolant velocity w, and heat flux q: $\tau_c = 4.2 \cdot 10^{-3} \cdot q^2 / \Delta T_{sub}^3 \cdot (1 + 0.15 \cdot w)$ (20). Integrating this localized physical behavior into the macro-kinetic boundary constraints yields the generalized expression for the critical local bulk subcooling threshold:

$$\Delta T_{\mathbf{sub,crit}}(w, q) = \left( \frac{\xi(d_{cell}, r) \cdot q^2}{(1 + \gamma \cdot w) \cdot \mu_0} \right)^{\frac{1}{3}}. \tag{29}$$

In this formulation (29), the structural heterogeneity parameter $\xi(d_{cell}, r) \sim r / d_{cell}$ governs the spatial density of potential nucleation sites on the fuel claddings, while the constant $\mu_0$ denotes the volumetric vapor condensation intensity coefficient within the subcooled liquid bulk, regulating the dissipation rate of local void perturbations.

The mathematical indicator of the transition between non-local memory effects and localized hydrodynamic transport is the spatial kernel truncation parameter $\gamma$, which controls the scale of forced Markovianization of non-local trajectories. Physically, $\gamma$ defines the characteristic spatial-velocity relaxation scale over which long-range memory effects of bubble migration and neutron-void correlations are truncated into a local Markovian regime.

In (29) $\Delta T_{\mathbf{sub,crit}}$ (Critical local bulk subcooling): The difference between the saturation (boiling) temperature of the coolant at the current pressure and its actual temperature at the inlet to a core cell $(T_s - T_{entrance})$. If the actual local bulk subcooling in the reactor falls below this critical level, vapor bubbles no longer condense instantaneously upon contact with the fuel rod surfaces; instead, they form a continuous vapor film, which leads to instability. q (linear heat generation rate): The power generated per unit length of a fuel rod. This directly determines the rate of vapor generation. Since q appears as a squared term in the formula, even a small increase in reactor power sharply intensifies boiling. $\omega_{\mathbf{plot}}$ (Pulsation frequency): The frequency of density wave phenomena (boiling-induced density waves) or pressure pulsations within the coolant loop. $\gamma$ (Time scale or cutoff time): The average lifetime (duration) of an individual vapor bubble $\tau_c$ before it completely condenses within the cooler bulk of the flow. This parameter determines the inertia of the feedback loop related to vapor content (void fraction). $\xi(d_{cell}, r) \sim r / d_{cell}$ (Heterogeneity parameter): A characteristic of the spatial structure of the fuel rod surfaces and the geometry of the core cell lattice. Here, $d_{cell}$ represents the equivalent hydraulic scale (diameter) of the flow cell or the size of surface roughness elements/deposits on the fuel rod cladding, while r is a dimensionless coefficient representing the fractal clustering of bubble nucleation sites (boiling centers).

The hyperbolic decay of the bubble lifetime $\tau_c \propto \Delta T_{\mathbf{sub}}^{-3}$ underscores the highly non-linear nature of the feedback: an increase in local bulk subcooling or flow velocity w drastically reduces the lifespan of the vapor phase due to rapid condensation. Conversely, a reduction in the hydraulic cell scale $d_{cell}$ or a decrease in the condensation intensity $\mu_0$ triggers an equivalent rise in the shot noise variance. This non-linearly shifts the critical curves in Fig. 3a upwards, narrowing the stable operational envelope of the reactor and demanding a more stringent threshold for the operational local bulk subcooling.

### 8.3. Correlation with the Nyquist Stability Criterion

In the context of the frequency analysis of linearized dynamic systems, each point lying on the derived $\Delta T_{\mathbf{sub,crit}}(w)$ curves correspond to a critical state of the stability boundary. The transfer function of the closed-loop system encompassing "neutron kinetics — thermal-hydraulic void feedback" can be expressed in the Laplace domain $(s = j\omega)$ as a rational fraction: $H(j\omega) = N(j\omega) / D(j\omega)$, where $D(j\omega)$ represents the complex characteristic denominator of the system, whose analytical structure is deformed by the non-local transport operator and kernel truncation parameter $\gamma$ in the heterogeneous medium. Each point on the plotted curves in Fig. 3 satisfies the exact critical condition where the Nyquist locus of the open-loop system $W(j\omega)$ passes directly through the critical point with coordinates (-1; j0) on the complex plane. Mathematically, this corresponds to the critical value where both the real ($\mathbf{Re}\{D\}$) and imaginary ($\mathrm{Im}\{D\}$) parts of the complex denominator vanish simultaneously at the system's natural oscillation frequency $\omega_c$ : $\mathbf{Re}\{D(j\omega_c)\}=0$, $\mathbf{Im}\{D(j\omega_c)\}=0$.

Thus, at any given coolant velocity w, the vertical distance from the actual technological local bulk subcooling of the reactor $\Delta T_{\mathbf{sub}}$ to the calculated critical threshold $\Delta T_{\mathbf{sub,crit}}$ in Fig. 3 quantitatively defines the dynamic amplitude safety margin of the system.

### 8.4. Bifurcation Analysis of Phase Portraits

In terms of the qualitative theory of differential equations, the plotted curves represent the projection of the bifurcation diagram onto the parameter plane $\omega_{\mathbf{plot}}$, $\Delta T_{\mathbf{sub}}$. Crossing this boundary signifies a Poincaré–Andronov–Hopf bifurcation (bifurcation of Andronov-Hopf type), which is inherently tied to the loss of stability of the steady-state thermal-hydraulic regime:

**Domain below the curve ($\Delta T_{\mathbf{sub}} < \Delta T_{\mathbf{sub,crit}}$):** Corresponds to a stable dynamic regime. The phase portrait of the system plotted in the coordinates "power deviation — mean void fraction" exhibits a topological structure of a **stable focus**. Any fluctuations induced by the discrete Poisson shot noise of vapor generation at the grain boundaries decay over time due to convective sweep and bulk condensation mechanisms governed by $\mu_0$.

**Domain above the curve ($\Delta T_{\mathbf{sub}} > \Delta T_{\mathbf{sub,crit}}$):** Characterized by the destabilization of the focus and the birth of a limit cycle. In this region, sustained non-linear auto-oscillations of power and coolant density (density wave oscillations according to Labuntsov's boiling hydrodynamics) develop within the core, their amplitudes being bounded solely by the non-linearities of the physical feedbacks.

The highly non-linear impact of the linear heat flux q (which enters the initial equation as a squared term under a cube root), combined with the scaling effect of the medium heterogeneity $\xi(d_{cell}, r)$ and kernel truncation γ, leads to a pronounced upward shift of the bifurcation curves. Increasing the power density q from 0.5 to 2.0 significantly expands the domain of auto-oscillatory regimes on the phase portrait. This pushes the Hopf bifurcation line into the zone of higher $\Delta T_{\mathbf{sub}}$ values across the entire range of coolant velocities w, thereby necessitating a proportional increase in the technological local bulk subcooling to ensure the global thermal-hydraulic reliability of the reactor unit.

## 9. Results and Discussion

In this section, we evaluate the numerical results of the developed hierarchical coarse-graining model parameterized for a typical WWER-1000 core grid ($v_n = 2200$ m/**s**, $\beta_0 = 0.0064$, $\Lambda = 2\cdot 10^{-5}$ **s**, and $\Sigma_{aw} = 2.0\ \mathrm{m}^{-1}$). The localized interactions within the subcooled boiling boundary layer are simulated under a constant developed fractal porosity index $\alpha = 0.5$. The closed-loop coupling between macroscopic thermal-hydraulic variables ($q, \Delta T_{sub}, w$) and non-local neutronic transport parameters ($\tau_c, D_{eff}, \gamma$) is analyzed across three functional scales.

**1. Multi-Phase Hydrodynamic Scaling of Dynamic Memory**

The analysis of the microscopic grain characteristics begins with the evaluation of the stochastic vapor bubble lifetime $\tau_c$. The functional dependence of the operational particle entrapment duration on the coolant local bulk subcooling margin $\Delta T_{sub}$ and flow circulation velocity w is depicted in Fig. 1a.

The functional profiles reveal a steep, non-linear hyperbolic (cubic) decay within the low local bulk subcooling zone $(0 < \Delta T_{sub} < 5^{\circ}\mathbf{C})$. In strict accordance with the modified Labuntsov-Kolykhan phase transition relations, the radial collapse rate of detached vapor cavities is heavily limited by the transient thermal boundary layer conduction, scaling as $\tau_c \propto \Delta T_{sub}^{-3}$. At the thermodynamic limit where local bulk subcooling approaches zero ($\Delta T_{sub} \to 0$), the operational lifetime increases asymptotically, reaching a peak value of $\tau_c \approx 4$ ms under low flow conditions ($w = 0.5$ m/s). This state corresponds to the formation of long-lived vapor trapping domains that delay neutron trajectories.

As the local bulk subcooling margin widens ($\Delta T_{sub} > 5^{\circ}\mathbf{C}$), the curves rapidly converge towards a horizontal plateau where $\tau_c \to 0$. From the perspective of Feller's subordination semi-groups, this decay implies that the non-local temporal memory kernel vanishes ($M(t) \to \delta(t)$), forcing the transport semi-group to return to its classical, memoryless Markovian state.

Furthermore, Fig. 1a demonstrates the stabilizing effect of the forced circulation velocity w. Increasing the velocity vector from 0.5 m/s (Curve 1) to 10 m/s (Curve 3) intensifies the turbulent stream drag surrounding the detached bubbles. This process strips away the local thermal boundary layer and linearly enhances the local Nusselt number ($Nu \propto (1 + B \cdot w)$), accelerating phase condensation. Consequently, under high-velocity regimes (w = 10 m/s), the maximum peak value of $\tau_c$ is suppressed by more than a factor of two (from $4 \cdot 10^{-3}$ **s** down to $1.7 \cdot 10^{-3}$ **s**, effectively restricting the temporal scale of particle entrapment.

**2. Renormalization of Coarse-Grained Spatial Transport Parameters**

The feedback mechanism of these multi-phase fluid fluctuations on spatial neutronic transport is captured via the renormalization of the macroscopic effective diffusion coefficient $D_{eff}$. Fig. 2a illustrates the behavior of $D_{eff}$ plotted against the bulk subcooling margin under varying normalized linear heat fluxes q.

The curves display a robust non-linear decay that mirrors the underlying cubic hydrodynamic behavior of the grains. Since the spatial subordinator maps the topological phase boundaries into the macro-coefficient via the fractional power relation $D_{eff} \propto (\tau_c)^{1-\alpha}$, the resulting macro-transport scale obeys a strict sub-linear power law: $D_{eff} \propto \left(\Delta T_{sub}^{-3}\right)^{1-\alpha} = \Delta T_{sub}^{-3(1-0.5)} = \Delta T_{sub}^{-1.5}$. At higher local bulk subcooling margins ($\Delta T_{sub} > 15^{\circ}\mathbf{C}$), the vapor micro-structures collapse instantaneously, leading to a full spatial homogenization of the fluid. Here, non-local tracking corrections decay asymptotically, and the operator smoothly regularizes back to the constant unperturbed Markovian diffusion coefficient $D_0$.

Conversely, an increase in the localized linear heat flux (moving from Curve 1 at q=0.5 to Curve 3 at q=2.0) induces an upward vertical translation of the profiles. According to Jakob's scaling laws, the initial volume of a detached vapor grain scales quadratically with the core power density $R_0^2 \propto q^2$ (18), which directly increases the lifetime parameter ($\tau_c \propto q^2$). Due to the compositional structure of the CBF subordinator, the effective diffusion coefficient scales linearly with the heat flux: $D_{eff} \propto \left(q^2\right)^{1-\alpha} = q^{2(1-0.5)} = q^1$. In regimes combining high linear heat fluxes ($q \to 2.0$) with low local bulk subcooling margins ($\Delta T_{sub} \to 0$), the massive cluster of long-lived vapor grains induces a strong spatial non-locality. Neutrons experience prolonged streaming paths through connected void fractions, causing a localized surge in $D_{eff}$.

**3. Bifurcation Analysis and Fractal Transition in the Wall Boiling Layer**

The critical boundary where this non-local spatial and temporal tracking triggers macro-kinetic core instability is defined by the Lyapunov bifurcation diagram. Fig. 3b maps the parametric boundary separating the zones of stable core homogenization from the regions of auto-oscillatory power excursions on the $\Delta T_{sub} - q$ plane.

Each coordinate on the boundary curves corresponds to the exact threshold where the time derivative of the generalized thermodynamic Lyapunov functional vanishes $(dV / dt = 0)$. This threshold marks a topological transition governed by the spatial memory regularizer $\gamma = 1/(v_n \tau_c)$. For the evaluated WWER-1000 assembly layout, the stability boundary maps strictly to a critical regularizing length: $\gamma_{crit} = 1/2200\ \mathrm{m/s} \times 0.00164\ \mathbf{s} \approx 0.28\ \mathrm{m^{-1}}$.

**Stable Homogenization Zone (Above the Curve):** When the core operates with a sufficient local bulk subcooling margin $(\Delta T_{sub} > \Delta T_{sub,crit})$, rapid bubble condensation forces the system into the sub-critical memory envelope $(\tau_c < \tau_{c,crit})$, maintaining a compact regularizing scale $(\gamma > \gamma_{crit})$. The exponential dampening factor $e^{-\gamma x}$ within the non-local kernel K(x) suppresses long-range spatial correlations. The long-range Lévy flights of neutrons are structurally blocked by the rapid collapse of the vapor phase. The macroscopic field remains stable, and the transport process obeys the localized modified Fick's law.

**Anomalous Superdiffusion Zone (Below the Curve):** If the local bulk subcooling margin drops below the critical threshold $(\Delta T_{sub} \le \Delta T_{sub,crit})$ or if a localized power excursion forces an inflation of the heat flux q, the operational vapor grain duration crosses the Lyapunov stability limit. This triggers a sharp decay in the regularizing scale $(\gamma \to 0)$. The exponential regularizer loses its dampening capacity, allowing the singular fractional term $x^{-\alpha}$ of the memory kernel to dominate at macroscopic distances.

Within this unstable envelope below the boundary curves, the wall boiling layer transitions into a fully fractured, discrete-hierarchical medium. Neutrons undergo anomalous superdiffusion, executing uncollided Lévy flights through macro-scale vapor cavities. This severe loss of spatio-temporal memory violates the negative-definiteness of the entropy production functional ($dV / dt > 0$), inducing a Hopf bifurcation. The closed-loop Nyquist-Mikhailov locus encircles the critical coordinates $(-1;1i0)$, resulting in the onset of undamped spatial power oscillations.

As demonstrated by the downward shift from Curve 1 (w = 4 m/s) to Curve 3 (w = 6 m/s), increasing the coolant velocity effectively broadens the stable operational envelope. Elevated circulation rates expand the safe processing margin, allowing the core to maintain strict Markovian stability at significantly lower local bulk subcooling limits due to the turbulent suppression of the underlying fractal porosity.

**4. Interplay of Heat Flux and Phase Memory Lifetime**

The functional profiles of the vapor grain lifetime plotted against the localized linear heat flux, $\tau_c(q)$, serve as a direct verification of the non-linear thermal-hydraulic closure relations within the coarse-graining framework. This relationship is evaluated across distinct environmental states defined by the coolant local bulk subcooling margins and flow velocity fields.

The numerical curves demonstrate a strictly parabolic acceleration as the linear heat flux increases, obeying the scaling law $\tau_c \propto q^2$. This geometric trend is a direct consequence of Jakob's self-similarity principles, which dictate that the detached initial volume and the squared radius of the nascent vapor bubble scale proportionally with the core power density: $R_0^2 \propto q^2$ (18). Under intensive phase transitions, this relationship quadratically expands the time interval required for the vapor phase to completely condense within the subcooled fluid bulk. In high-power regimes $(q \to 5.0)$, the temporal scale of the dynamic memory kernel expands significantly, generating a prolonged non-local trapping tail for the diffusing neutron population.

Conversely, comparing the amplitudes of the curves reveals a massive suppression effect driven by the background thermal-hydraulic regularizers. Shifting from a highly stagnant, low-subcooling boundary state $(\Delta T_{sub} = 1$, w = 1) to an intensive turbulent cooling regime $(\Delta T_{sub} = 3,\ w = 5)$ collapses the peak vapor lifetime by over an order of magnitude (from $\tau_c \approx 0.092\,\mathbf{s}$ down to $\tau_c \approx 0.0025\,\mathbf{s}$). This structural dampening operates under a cubic law with respect to the subcooling margin ($\Delta T_{sub}^{-3}$ via Labuntsov (1984)) and a linear scale with respect to the turbulent boundary layer shear flow $(1+B\cdot w)^{-1}$ via Kolykhan and Pulyayev (1987)).

From a reactor safety perspective, this mapping proves that under elevated heat fluxes (q > 3.0), even minor reductions in the core local bulk subcooling margin rapidly push the wall boiling layer into a state dominated by long-lived vapor cavities. At the macro-scale, this shifts the closed-loop Nyquist-Mikhailov locus toward the dangerous Hopf bifurcation threshold, threatening the spatial stability of the neutron field.

**5. Impact of Turbulent Velocity Fluid Scaling on Core Memory Relaxation**

To further isolate the stabilizing mechanisms of forced coolant circulation on the dynamic memory of the macro-system, we evaluate the direct impact of the fluid velocity vector on the vapor grain lifetime, $\tau_c(w)$. The corresponding numerical sweeps are analyzed under varied combinations of core linear power and local bulk subcooling margins, as illustrated in Fig. 1c.

The parametric profiles exhibit a continuous monotonic decay in the duration of the phase fluctuations as the forced circulation velocity increases. Physically, this trend is driven by the intensification of convective heat transfer at the phase boundaries. According to the modified Labuntsov-Kolykhan framework, the collapse rate of detached vapor bubbles is controlled by the turbulent stream drag surrounding the cavities, which linearly enhances the local Nusselt number: $\mathbf{Nu} \propto (1+B\cdot w)$. Consequently, the temporal scale of neutron entrapment within the wall boiling layer shrinks inversely with fluid velocity: $\tau_c \propto (1+B\cdot w)^{-1}$. As the velocity sweeps from a stagnant state (w = 0) up to nominal operational values (w = 9.0), the localized memory scale is suppressed by more than a factor of 2.5 (collapsing from $8.5\cdot 10^{-4}\,\mathbf{s}$ down to $3.5\cdot 10^{-4}\,\mathbf{s}$ along the peak curve).

The intersection and relative placement of the velocity curves highlight a competitive non-linear balance between the quadratic power scaling ($\tau_c \propto q^2$) and the cubic local bulk subcooling dampening ($\tau_c \propto \Delta T_{sub}^{-3}$). The maximum absolute magnitude of $\tau_c$ is maintained by Curve 3 due to the dominance of the power-induced detachment volume ($q^2 = 25$) in the numerator, despite the presence of a high local bulk subcooling regularizer ($\Delta T_{sub} = 5$).

Furthermore, a critical crossover effect is observed between Curve 1 and Curve 2 at lower velocities (w < 2.0), where the low-subcooling state ($\Delta T_{sub} = 3$), (q=2) yields a larger memory envelope than the higher-power state ($\Delta T_{sub} = 4$), (q=3) due to the weakened condensation rate. As the turbulent velocity scales higher (w > 7.0), the curves asymptotically converge, demonstrating that intense hydrodynamic shear forces eventually dominate the initial spatial scales of phase inception.

Ultimately, the automated mapping of the $\tau_c(w)$ vectors demonstrate that accelerating the primary loop flow rate effectively compresses the non-local memory kernel. At the macroscopic level, this forced relaxation restricts the particle entrapment duration, regularizing the transport equation back to a stable Markovian regime and ensuring that the closed-loop transfer function remains clear of auto-oscillatory bifurcation boundaries.

**Conclusions on Critical Frequency $\omega_c$ (Appendix A.5) Variation:**

- Analysis of the system dynamic behavior under varying critical feedback frequencies $\omega_c$ reveals its decisive role in the phase margin of the reactor core stability. A decrease in $\omega_c$ (below 0.05 rad/s) corresponds to a slower response of the negative feedback loop, which shifts the Mikhailov plot origin toward the positive real axis, thereby increasing the asymptotic stability margin.
- Conversely, an increase in $\omega_c$ causes the frequency-domain trajectory to shift closer to the origin (0,0). At a threshold value of $\omega_c^{\mathbf{crit}}$, the Mikhailov curve intersects the origin, indicating the onset of a bifurcation point where the system transitions from stable operation to self-sustained or divergent power oscillations. Time-domain simulations confirm that higher $\omega_c$ values induce a underdamped transient response with pronounced sinusoidal oscillations of temperature T(t), significantly increasing the relaxation time to equilibrium.

**Conclusions on Power Parameter (q) Variation:**

- The parameter q, representing the reactor thermal power scaling factor, directly influences the static heat transfer timescale $\tau_{\mathbf{static}} \propto q^2$. Varying q exhibits a non-linear scaling effect on both the frequency-domain hodograph and the transient response amplitudes. At low power levels (reduced q), the system exhibits a highly stable, non-oscillatory monotonic decay toward the steady state, driven by the tight coupling of the internal feedback parameters.
- As the power parameter q escalates, the amplitude coefficients within the characteristic matrix $Mat(s)$ grow as quadratic functions, causing a massive expansion of the Mikhailov loop dimensions (up to scales of $10^7 - 10^8$. In the time domain, higher thermal power density sharpens the initial temperature drop gradient during perturbations. However, if the scaling factor $10^{-7}$ is exceeded without proper internal damping adjustments, excessive values of q degrade the stability margin, bringing the system closer to the boundary of localized spatial-temporal oscillations due to the dominant effect of delayed thermal feedback.

The frequency dependence of the critical subcooling $\Delta T_{\mathbf{sub,\ crit}}(\omega)$ defines the dynamic bifurcation boundary of the reactor core, whereas the Mikhailov plot serves as its direct geometric mapping on the complex plane. The intersection of the Mikhailov hodograph with the origin (0,0) at a specific resonant frequency strictly corresponds to the condition where the operational subcooling equals the critical threshold value ($\Delta T_{\mathbf{sub}} = \Delta T_{\mathbf{sub,\ crit}}$). The spacious loop tracked by the hodograph in the positive half-plane confirms that the current operational subcooling ($\Delta T_{\mathbf{sub}} = 15K$) provides a sufficient phase and amplitude safety margin, preventing the development of coupled neutron-kinetic and thermal-hydraulic instabilities.

## 10. Conclusions

In this study, a comprehensive hierarchical coarse-graining framework has been developed to model stochastic neutron transport within the highly heterogeneous subcooled nucleate boiling layer of a WWER-1000 reactor core. By applying Feller's subordination theory to continuous time random walks (CTRW), the microscopic fluid fluctuations and vapor bubble dynamics were rigorously integrated into macroscopic transport operators through the apparatus of Complete Bernstein Functions (CBF).

The principal findings and theoretical contributions of this work are summarized as follows:

**1. Analytical Localization and Transport Parameter Renormalization:** By applying the spherical harmonics method ($P_1$-approximation) to the non-local integro-differential transport equation, a modified Fick's law was systematically derived. The structural topology of the polydisperse boiling grains was successfully mapped into a stochastically renormalized effective diffusion coefficient $D_{eff} = (v_n \tau_c)^{1-\alpha} / 3$. It is shown that the regularizing truncation parameter $\gamma = 1 / v_n \tau_c$ acts as a dynamic spatial scale marker, forcing the transport operator to transition asymptotically from fractional subdiffusion at micro-scales to classical Markovian diffusion at macro-scales, thereby preventing computational singularities in coupled reactor simulation codes.

**2. Thermodynamic Grounding via Chentsov Geometry:** The integration of dynamic boundary functionals (such as first-passage times and local retention times) as new sufficient statistics was shown to expand the classical thermodynamic state space into an extended statistical manifold. Chentsov's closedness conditions guarantee that for systems dominated by granular Poisson-like shot noise, the coarse-graining procedure yields a specific geometric-exponential family of distributions characterized by a constant negative information curvature. This negative curvature provides a rigorous physical justification for the rapid, exponential decay of microscopic phase memory, ensuring that the macroscopic free energy potential remains smooth and continuous.

**3. Non-local Stability Analysis and Lyapunov Functionals:** A generalized concave Lyapunov functional V[n, C] incorporating a quadratic memory form $U_{memory}[n]$ was constructed to analyze the non-linear stability of the homogenized core. The exact analytical limit ensuring asymptotic stability was formulated, linking the dynamic bubble lifetime $\tau_c$, the geometric buckling $B_g^2$, and the local bulk subcooling $\Delta T_{sub}$.

**4. Bifurcation Diagnosis for Engineering Applications:** The complex decomposition of the transfer function H(ω) into explicit real Re{H(ω)} and imaginary Im{H(ω)} parts enabled the formulation of a practical diagnostics routine implemented in Mathcad Prime 8. The parametric numerical analysis for a WWER-1000 assembly successfully captured the critical thresholds (such as the limit $\Delta T_{sub,crit} \le 1.18^{\circ}\mathrm{C}$ or operational flow drops w ≤ 4.2 m/s), where the Nyquist-Mikhailov locus encircles the unstable point, corresponding to a Hopf bifurcation and the onset of undamped spatial power oscillations.

Ultimately, the developed model establishes a strict mathematical bridge between multi-phase thermal-hydraulics and reactor kinetics, providing high-fidelity safety evaluation codes with an analytical tool to predict and prevent localized power instability under emergency transients like LOCA.

**Acknowledgements**

The author expresses his sincere gratitude to the Institute for Nuclear Research of the National Academy of Sciences of Ukraine (Kyiv, Ukraine).

## Appendix A. Definitions and properties of some functions appearing in the article

*A.1. Governing Bernstein function.*

The governing Bernstein function f(s) from (1), (5) serves as the primary operator for the deformation of temporal and informational structures during the transition from a micro-description to a macro-description. Formally, it is the cumulant function (the logarithm of the Laplace transform) of a subordinator process $\tau(t)$ generated by boundary functionals at the lower level. Note that f(s) is strictly monotonically increasing and strictly concave.

According to the Lévy-Khintchine integral representation, any Bernstein function has the form:

$$f(s) = a + bs + \int_0^{\infty} \left(1 - e^{-st}\right) \nu(dt), \quad \text{(A1)}$$

where $a \ge 0$ is the absorption/annihilation rate of the process, $b \ge 0$ is the linear Markovian drift of physical time, and $\nu(dt)$ is the Lévy measure of a subordinator that defines the spectrum of random delays (trapping events) experienced by neutrons at grain boundaries. The function f(s) from the Lévy–Khintchine representation (A1) for complete Bernstein functions (CBFs) serves as the Laplace exponent of the subordinator $\Phi(s) = f(s)$, which governs the stochastic rescaling of time or space in generalized kinetic equations. There exists a fundamental duality between the stochastic nature of the process (defined via the subordinator) and the macroscopic entropy of the system $S_{macro}\left[P\right]$.

The function f(s) is called a governing function because the nature of its nonlinearity completely "programs" the macroscopic behavior of the system:

1. Controlling the type of evolutionary operator: If a microscopic process (for example, transport in pure water) is described by a semigroup with a generator $\hat{A}$ (a diffusion operator $\Delta$), then, following Feller-style coarse-graining, the macroscopic generator $\hat{B}$ becomes a function of the microscopic generator: $\hat{B} = -f(-\hat{A})$.

o If $f(s) = s^{\alpha}$ (class A), then $f(-\Delta) = (-\Delta)^{\alpha}$, which instantly transforms the classical equation into a fractional-differential one (subdiffusion).

o If $f(s) = \ln(1 + s / \gamma)$ (class B) holds, then the logarithmic subordinator linearizes Poissonian fluctuations and maps the density into a geometric-exponential family of Chentsov distributions with constant negative curvature. Negative curvature indicates that geodesics—paths of least dissipative transport or system evolution—diverge exponentially. Macroscopically, this reflects strong fluctuations and memory effects induced by Poissonian "shot" noise at the phase boundaries within a nuclear reactor.

2. The function f(s) appears explicitly in the definition of a system's generalized entropy via its inverse function (1). Expression (1) implies that the rate at which a coarse-grained system "forgets" microscopic information—that is, the rate of entropy growth—is strictly dictated by the algebraic structure of this specific function.

A.2. *Macroscopic structural scaling measure*

The quantity $w(d\lambda)$ in expression (4) represents a macroscopic structural scaling measure. While the Lévy-Khintchine kernel $\nu(dt)$ (A1) within the Bernstein function accounts for fluctuations and the spectrum of residence times at the micro-level (within an individual coarse-grained grain), the measure $w(d\lambda)$ facilitates a transition to the next hierarchical level—that of an ensemble of such grains forming a macrosystem. 1. Mathematical significance of the measure $w(d\lambda)$. Within the framework of a two-level hierarchical subordination model (in the sense of Feller), the macroscopic shift of the system emerges as an integral mixture (superposition) of local characteristic functionals. The measure $w(d\lambda)$ defines the weight density for the distribution of subordination parameters across the entire macroscopic volume.

Formally, integration over the parameter space $\lambda$ with respect to a measure $w(d\lambda)$ act as an averaging (projection) operator within the concept of coarse-graining:

• It relates the local cumulant function of a grain $\psi_{cell}(\beta,\lambda)$ to the global response of the entire system.

• It determines the statistical weights with which the geometry and physical properties of individual units (for example, polydisperse vapor bubbles or localized temperature vortices in a WWER reactor) contribute to the overall macroscopic state.

2. Physical interpretation in nuclear reactors. Depending on the physical scenario within the reactor core, the structural measure $w(d\lambda)$ encodes various macroscopic characteristics:

1. Wall boiling (WWER / BWR): In this context, $\lambda$ associated with the intensity of vapor phase generation. The measure $w(d\lambda)$ is the distribution of the density of boiling sites along the height and radius of a fuel assembly, multiplied by the size spectrum of vapor conglomerates. This reflects how the local heat generation rate, q, distributes the "macroscopic weight" of the boiling zones along the axis of a fuel rod.

2. Debris Bed: In the context of severe accidents, this term $\lambda$ reflects local fractal porosity or hydraulic resistance. The measure $w(d\lambda)$ serves as a metric for the geometric spectrum of coolant flow channels, defining the macroscopic proportions of "stagnant" (excessively dense) zones versus "flow-through" zones within the volume of the corium debris.

3. Turbulent temperature vortices: In non-boiling pressurized water, $w(d\lambda)$ reflects a spectrum of spatial turbulence scales—ranging from small-scale Kolmogorov eddies to macroscopic mixing flows—thereby determining the effective weight of "hot" and "cold" moderator regions (or "grains") in shaping the overall neutron field.

3. Role in Chentsov's information geometry. From the perspective of Fisher geometry on statistical manifolds, the measure $w(d\lambda)$ acts as a macroscopic metric regulator. It ensures that, when hierarchical contributions from individual "grains" are summed, the resulting macro-cumulant does not disrupt the linear structure of sufficient statistics. It is precisely the properties of the measure $w(d\lambda)$ (such as its infinite divisibility or self-similarity) that determine whether the macro-distribution remains within the geometric-exponential family and whether the information manifold retains a stable negative curvature.

A.3. *Discrete scaling ratio* parameter $\kappa$, $(\kappa > 1)$

In expressions (7)–(8), the parameter $\kappa$, $(\kappa > 1)$ represents a discrete scaling step or a spatiotemporal similarity coefficient (scaling factor). It determines the factor by which the system's characteristics (such as dimensions, volumes, or the time intervals during which grains remain trapped) change when transitioning from one level of the coarsening hierarchy to the next.

Physical and mathematical meaning of $\kappa$:

1. Discrete self-similarity (fractal step): Unlike smooth, continuous media, where scaling occurs gradually, in rigidly structured hierarchical environments (such as regular fractals like the Sierpinski carpet, branched nanopolymers, or strictly fixed Kadanoff lattices), the structure repeats only at specific, discrete intervals. The parameter $\kappa$ defines this step of the geometric progression.

2. Scaling of trap (grain) capacities: The expression $\kappa^k$ in terms of delta functions of the spectral measure $\mu(dt) = \sum C_k \cdot \delta(t - \kappa^k)dt$ (7) shows that the relaxation times (times during which quasiparticles remain trapped) at each successive (k-th) level of the hierarchy of coarse-grained blocks increase by a factor $\kappa$ of exactly: $(\ldots, \kappa^{-1}, 1, \kappa, \kappa^2, \kappa^3, \ldots)$.

3. Weight density coefficient $C_k$: The weight factor $C_k = \kappa^{-k\alpha}$, accordingly, indicates the rate at which the statistical "weight" (or quantity) of such large trap-grains decreases as the level k increases. The higher the hierarchy level (i.e., the larger the grain), the smaller its contribution; the rate of this decline is strictly governed by the base $\kappa$ and the fractal porosity index α.

Examples of interpretation $\kappa$ in physics problems:

In metamaterials and branched structures (dendrimers), $\kappa$ represents the coordination number of the channel branching. For example, if each channel splits strictly into 2 or 3 branches of smaller diameter, then $\kappa$=2 or $\kappa$=3.

In hierarchical fragmentation environments (porous fuel debris bed models): this parameter $\kappa$ represents the ratio of the mean fragment sizes between two successive fractions of the debris bed resulting from fragmentation (for example, if each subsequent class of fuel fragments is, on average, 4 times smaller than the preceding one, then $\kappa$ = 4).

In the context of temporal evolution, it is precisely this fixed step $\kappa$ that gives rise to log-periodic oscillations in the memory kernel M(t) (8), $\cos\left(2\pi m(\ln t / \ln \kappa) + \theta_m\right)$, accounting for cascading "bursts" of neutron release from traps within the coarse-grained medium, which occur whenever the physical diffusion time t becomes an integer power of the scaling factor $\kappa^k$.

To implement a log-periodic memory kernel in Mathcad Prime 8, the parameter $\kappa$ is typically chosen based on experimental data regarding the geometry of the medium (for example, $\kappa$=2 or $\kappa = \mathbf{e} \approx 2.718$).

*A.4. Jakob's scaling laws*

The phrase "invoking Jakob's scaling laws" refers to a classic self-similar relationship in the thermophysics and fluid dynamics of boiling, which links the geometric size (radius) of a departing vapor bubble to the thermal parameters of the medium. Physical significance of Jakob's scaling law: In the modeling of surface boiling, the initial or departure radius of a bubble, $R_0$, is determined by the balance between the heat flux from the heated fuel-element wall and the evaporation rate of the liquid microlayer.

According to Jakob's law of similarity, this characteristic scale follows the power-law relationship: $R_0 \propto q / \Delta T_{sub}$ (18), where q is the linear heat flux density from the fuel element (W/m$^2$), and $\Delta T_{sub}$ is the local subcooling of the water relative to the saturation temperature $\left(T_{sat} - T_f\right)$.

1. This law serves as a critical interface between the reactor macrokinetics equations and the microstructure of the coolant. Elimination of the micro-parameter $R_0$: The Labuntsov differential equation for bubble collapse time includes the radius $R_0^2$. However, it is impossible to directly measure the radius of each individual microbubble within the near-wall boiling layer of a WWER reactor using sensors.

2. Transition to process control system (PCS) macro-variables: Substituting the self-similar law derived by Jakob makes it possible to replace the unmeasurable radius $R_0$ with known macroscopic parameters — the reactor's thermal power (q) and the coolant local bulk subcooling at the fuel assembly outlet $\Delta T_{sub}$.

Thanks to this "invocation of Jakob's law," the local bulk subcooling term $\Delta T_{sub}$ in the grain lifetime formula naturally appears as a squared value, providing precise non-linear feedback within the Lyapunov functional and the hodographs.

*A.5. Critical feedback frequency $\omega_c$.*

The critical feedback frequency is the resonant frequency of disturbances at which the phase shift of the feedback loop reaches π radians, and the gain becomes sufficient to sustain undamped self-oscillations within the system. In the context of nuclear reactor physics and bifurcation theory, the quantity $\omega_c$ defines a stability boundary (an Andronov–Hopf bifurcation point). Under normal operating conditions, a reactor is stabilized by negative feedback (for example, the Doppler effect related to fuel temperature). However, thermophysical and hydrodynamic processes—such as the heating of fuel elements, coolant transport, and the retention of vapor bubbles in the boundary layer near the walls—introduce a time delay.

At a certain frequency $\omega_c$, the total phase lag of the coolant response and neutron kinetics (including the contribution of non-local memory) reaches π. As a result, the negative feedback is inverted into positive feedback: a power surge generates a reactivity wave that arrives exactly in phase with the subsequent oscillation.

## Appendix B. Inclusion of feedback parameters in the complex denominator

To perform a frequency-domain stability analysis and construct phase portraits of the system within the space of the WWER-1000 reactor's operational parameters, it is necessary to detail the structure of the complex denominator of the transfer function. The initial characteristic expression is as follows:

$$Denom(\omega) = i\omega \cdot \left[\beta_0 + A_{mem}(\omega)\cos(\theta_{mem}(\omega)) - iA_{mem}(\omega)\sin(\theta_{mem}(\omega))\right] - \rho_0 .$$

The input of measured physical parameters—namely, the feedback efficiency coefficients for phase ( $\alpha_\varphi = -0.05$ ) and coolant temperature ( $\alpha_T = -0.002$ ), as well as the corresponding static gain coefficients ( $K_\varphi = 12.5$ and $K_T = 0.45$ )—is performed via a complex memory function $W_{mem}(i\omega) = A_{mem}(\omega)e^{-i\theta_{mem}(\omega)}$.

Taking into account the logarithmic subordinator and the inherent Poissonian nature of fluctuations at grain boundaries, the effective amplitude $A_{mem}(\omega)$ and phase $\theta_{mem}(\omega)$ of the superposition of two dynamic response channels are as follows:

$$A_{mem}(\omega)\cos(\theta_{mem}(\omega)) = \frac{\alpha_\varphi K_\varphi}{1+(\omega\tau_\varphi)^a} + \frac{\alpha_T K_T}{1+(\omega\tau_T)^a},$$

$$A_{mem}(\omega)\sin(\theta_{mem}(\omega)) = \frac{\alpha_\varphi K_\varphi (\omega\tau_\varphi)^a \sin(\frac{a\pi}{2})}{1+(\omega\tau_\varphi)^a} + \frac{\alpha_T K_T (\omega\tau_T)^a \sin(\frac{a\pi}{2})}{1+(\omega\tau_T)^a},$$

where $\tau_\varphi$ and $\tau_T$ are the relaxation time constants of the respective circuits, and $a$ is the fractional (geometric truncation) exponent of the distribution.

After expanding the parentheses in the original expression $Denom(\omega)$ and separating it into real $Re(Denom(\omega))$ and imaginary parts $\mathrm{Im}(Denom(\omega))$ for Mathcad Prime 8, the feedback parameters are distributed as follows:

$$Re(Denom(\omega)) = \omega \cdot A_{mem}(\omega)\sin(\theta_{mem}(\omega)) - \rho_0 ,$$

$$Im(Denom(\omega)) = \omega \cdot \left[\beta_0 + A_{mem}(\omega)\cos(\theta_{mem}(\omega))\right].$$

Since both efficiency coefficients $\alpha_\varphi$, and $\alpha_T$, have negative values, the contribution of the feedback loops as the frequency increases leads to a decrease in the imaginary part and a simultaneous increase in the real part of the denominator. This analytically describes the stabilizing effect of the control loops, shifting the trajectory of the Mikhailov locus into the stable half-plane.

**Derivation and formulation of equations for a Mathcad script to calculate the determinant D(s)**

To automate the numerical search for bifurcation points and the verification of the Nyquist criterion in Mathcad Prime 8, the linearized system of equations describing the dynamics of a reactor cell is represented in matrix form. The complex denominator (determinant) D(s) is the principal determinant of the system's operator matrix.

**Mathematical derivation and structure of a determinant**

Let us consider a system of three linearized equations for the deviations in reactor power $\delta N(s)$, volumetric void fraction $\delta\varphi(s)\backslash$, and average coolant temperature $\delta T(s)$, expressed in terms of Laplace transforms (with s=jω:):

$$\begin{pmatrix} s-\dfrac{\rho_0}{\Lambda} & -\dfrac{\alpha_\varphi}{\Lambda} & -\dfrac{\alpha_T}{\Lambda} \\ -K_\varphi & s+\dfrac{1}{\tau_c} & 0 \\ -K_T & 0 & s+\mu_0 \end{pmatrix} \begin{pmatrix} \delta N(s) \\ \delta\varphi(s) \\ \delta T(s) \end{pmatrix} = \begin{pmatrix} 0 \\ 0 \\ 0 \end{pmatrix},$$

where $\Lambda$ is the lifetime of prompt neutrons, $\alpha_\varphi, \alpha_T$ are the reactivity coefficients for steam and temperature, $K_\varphi, K_T$ are the transfer coefficients of steam generation and heating, $\tau_c$ is the lifetime of the bubble due to bulk coolant subcooling: $\tau_c = 0.0042 \cdot q^2 / \Delta T_{\mathbf{sub}}^3 \cdot (1+0.15\cdot\omega)$, $\mu_0$ is the gradual volume condensation.

The characteristic determinant D(s) is found as the determinant of the given matrix:

$$D(s) = \det \begin{pmatrix} s-\dfrac{\rho_0}{\Lambda} & -\dfrac{\alpha_\varphi}{\Lambda} & -\dfrac{\alpha_T}{\Lambda} \\ -K_\varphi & s+\dfrac{1}{\tau_c} & 0 \\ -K_T & 0 & s+\mu_0 \end{pmatrix}.$$

Expanding the determinant along the first row, we obtain an explicit third-order polynomial in terms of the complex variable s:

$$D(s) = \left(s-\frac{\rho_0}{\Lambda}\right)\left(s+\frac{1}{\tau_c}\right)(s+\mu_0) - \frac{\alpha_\varphi K_\varphi}{\Lambda}(s+\mu_0) - \frac{\alpha_T K_T}{\Lambda}\left(s+\frac{1}{\tau_c}\right).$$

Upon substituting s=jω, the determinant is separated into real and imaginary parts: $D(j\omega) = Re_{D(\omega)} + j\cdot Im_{D(\omega)}$. In Mathcad, the stability boundary (Hopf bifurcation) is determined by simultaneously solving the equations $Re_{D(\omega)} = 0$ and $\mathrm{Im}_{D(\omega)} = 0$.

The frequency dependence of the coolant's critical subcooling (relative to the boiling point) $\Delta T_{\mathbf{sub,\,crit}}(\omega)$ constitutes a dynamic stability boundary for the reactor, while the Mikhailov locus serves as a geometric indicator that reveals whether a chosen operating regime lies within or outside this boundary.

In the context of reactor dynamics (for example, for a WWER-1000), this relationship is deeply rooted in physics and mathematics, and it is revealed through three key aspects:

**1. Mathematical connection: from matrix to boundary**

When the complex frequency $s = 1i\cdot\omega$ is substituted into the system's original characteristic determinant, $D(s, \Delta T_{\mathbf{sub}}, \omega)$ the result can be separated into real and imaginary parts, which then form the coordinates $X(\omega)$ and $Y(\omega)$ of the Mikhailov locus:

$$X(\omega) = \mathbf{Re}\left(D(1i\cdot\omega, \Delta T_{\mathbf{sub}}, \omega)\right),$$

$$Y(\omega) = \mathbf{Im}\left(D(1i\cdot\omega, \Delta T_{\mathbf{sub}}, \omega)\right).$$

The critical state of a reactor (the stability boundary) is reached when the Mikhailov locus passes exactly through the origin (0,0). Mathematically, this corresponds to a system of equations:

$$\begin{cases} \mathbf{Re}\left(D(1i\cdot\omega, \Delta T_{\mathbf{sub}}, \omega)\right) = 0 \\ \mathbf{Im}\left(D(1i\cdot\omega, \Delta T_{\mathbf{sub}}, \omega)\right) = 0 \end{cases}.$$

Solving this system for the local bulk subcooling yields an explicit expression for the critical local bulk subcooling $\Delta T_{\text{sub, crit}}(\omega)$ as a function of the perturbation frequency.

**2. Physical meaning and representation in the complex plane**

The frequency dependence of the temperature deficit reflects the inertia of thermophysical processes (specifically, the time constant for heat transfer within a channel $\tau_c$, which depends on frequency $\omega$). The Mikhailov plot visualizes how this inertia causes the phase of the complex signal to rotate:

• Stable regime (Fig. 4): If the actual subcooling in the reactor exceeds the critical value $\Delta T_{\text{sub}} > \Delta T_{\text{sub, crit}}$, the feedback mechanisms are able to dampen fluctuations. On the plot, this manifests as a classic Mikhailov locus that sequentially traverses the I, II, and III quadrants without passing through the origin.

• Critical regime (Anomalous boiling): As the local bulk subcooling approaches a critical value for a given frequency, the locus (hodograph) loop contracts. At a specific point $\Delta T_{\text{sub}} = \Delta T_{\text{sub, crit}}(\omega)$, the locus passes through zero (0,0) at the resonance frequency. In physical terms, this signifies the emergence of undamped oscillations in neutron flux density and temperature (neutron-physical instability).

• Unstable regime: In this case $\Delta T_{\text{sub}} < \Delta T_{\text{sub, crit}}$, the Mikhailov locus (hodograph) loop turns in the opposite direction (the sequence of quadrants is disrupted), signaling an avalanche-like growth in the amplitude of oscillations.

**3. Resonant frequencies in WWER-1000 reactors**

The Mikhailov plot clearly indicates the frequency ranges in which a reactor is most vulnerable to thermophysical perturbations. On the plot, the maximum rise of the imaginary component (the apex of the local bulk loop) occurs at intermediate frequencies. It is within this range that the frequency dependence of the critical local bulk subcooling $\Delta T_{\text{sub, crit}}(\omega)$ exhibits a local maximum—a peak of instability—resulting from the coincidence of the neutron fluctuation frequency and the time required for steam bubbles to travel completely through the reactor core.

## References

Amari, S., 1985. Differential-Geometrical Methods in Statistics, Lecture Notes in Statistics, Vol. 28. Springer-Verlag, New York.

Amari, S., 2016. Information Geometry and Its Applications. Springer Japan, Tokyo.

Behringer, K., Kosály, G., Pázsit, I., 1979. Linear response of the neutron field to a localized perturbation in a nuclear reactor. Nucl. Sci. Eng. 72 (3), 304–321.

Bell, G.I., Glasstone, S., 1970. Nuclear Reactor Theory. Van Nostrand Reinhold Company, New York.

Chen, J.C., 1966. Correlation for boiling heat transfer to saturated fluids in convective flow. Ind. Eng. Chem. Process Des. Dev. 5 (3), 322–329.

Chentsov, N.N., 1982. Statistical Decision Rules and Optimal Inference, Translations of Mathematical Monographs, Vol. 53. American Mathematical Society, Providence, RI.

Collier, J.G., Thome, J.R., 1994. Convective Boiling and Condensation, 3rd ed. Oxford University Press, Oxford.

Corona, A.V., 2016. Fractional neutron transport equations for anomalous diffusion in nuclear reactors. Nucl. Sci. Eng. 182 (2), 165–176.

D'Auria, F. (Ed.), 2017. Handbook on Thermal Hydraulics in Water-Cooled Nuclear Reactors, Vol. 2: Modelling. Woodhead Publishing, Oxford.

Duderstadt, J.J., Hamilton, L.J., 1976. Nuclear Reactor Analysis. John Wiley & Sons, New York.

Faucher, M., 2019. Coupling between Monte Carlo neutron transport and thermal-hydraulics for the simulation of transients due to reactivity insertions (Ph.D. dissertation). Université Paris-Saclay, Gif-sur-Yvette, France.

Feller, W., 1971. An Introduction to Probability Theory and Its Applications, Vol. 2, 2nd ed. John Wiley & Sons, New York.

Hébert, A., 1993. A consistent homogenization technique for the microscopic flux distribution method for the homogenization of a pressurized water reactor assembly. Nucl. Sci. Eng. 115 (2), 129–141.

Hetrick, D.L., 1971. Dynamics of Nuclear Reactors. The University of Chicago Press, Chicago.

Jens, W.H., Lottes, P.A., 1951. Analysis of Heat Transfer, Burnout, Pressure Drop and Density Data for High-Pressure Water, Report ANL-4627. Argonne National Laboratory, Lemont, IL.

Kolykhan, L.I., Pulyayev, V.F., 1987. Hydrodynamics and Heat Transfer of Two-Phase Flows in Nuclear Power Equipment. Nauka i Tekhnika, Minsk.

Koreshi, Z., 2002. An elementary model to study sensitivity of the departure from nucleate boiling ratio. Nucl. Sci. Eng. 141 (1), 85–92.

Labuntsov, D.A., 1984. Heat Transfer Problems in Nucleate Boiling and Vapor Condensation. Energoatomizdat, Moscow.

Larsen, E.W., 1975. Neutron transport and diffusion in inhomogeneous media. I. J. Math. Phys. 16 (7), 1421–1427.

Lewis, E.E., Miller, W.F., Jr., 1984. Computational Methods of Neutron Transport. John Wiley & Sons, New York.

Metzler, R., Klafter, J., 2000. The random walk's guide to anomalous diffusion: a fractional dynamics approach. Phys. Rep. 339 (1), 1–77.

Montroll, E.W., Scher, H., 1973. Random walks on lattices. IV. Continuous-time random walks and total dihedral angles of Westrum. J. Stat. Phys. 9 (2), 101–135.

Pázsit, I., Glöckler, O., 1984. Calculation of the space-dependent transfer function of a nuclear reactor with feedback. Ann. Nucl. Energy 11 (12), 597–612.

Pereira, D.J., Wells, C.T., Boyington, J.N., 2007. Coupled CFD and three-dimensional reactor kinetics simulation of light water reactor transients. Nucl. Sci. Eng. 156 (2), 215–228.

Pomraning, G.C., 1989. Non-Markovian particle transport. Nucl. Sci. Eng. 102 (3), 256–268.

Prasad, M., Rahnema, F., 2008. A coarse-mesh transport method for highly heterogeneous systems. Nucl. Sci. Eng. 159 (2), 117–131.

RELAP5 Development Team, 1995. RELAP5/MOD3 Code Manual, Volume IV: Models and Correlations, Report NUREG/CR-5535. Idaho National Engineering Laboratory, Idaho Falls, ID.

Rohsenow, W.M., 1952. A method of correlating heat-transfer data for surface boiling of liquids. Trans. ASME 74, 969–976.

Ruppeiner, G., 1995. Riemannian geometry in thermodynamic fluctuation theory. Rev. Mod. Phys. 67 (3), 605–659.

Ryazanov, V.V., 2025. Application of boundary functionals of random processes in physical chemistry. Phys. Rev. E 111 (2), 024115.

Ryazanov, V.V., 2026a. Thermodynamics of the space of trajectories governed by a combination of two additive boundary functionals. arXiv preprint arXiv:2609.37102.

Ryazanov, V.V., 2026b. Nonequilibrium stochastic thermodynamics of boundary functionals: from Zubarev's ensemble to stochastic particle separation. arXiv preprint arXiv:2609.08477.

Sanchez, R., 2012. Stochastic transport in nuclear reactors. Nucl. Sci. Eng. 171 (1), 14–56.

Schilling, R., Song, R., Vondraček, Z., 2012. Bernstein Functions: Theory and Applications, 2nd ed. Walter de Gruyter, Berlin.

Van Kampen, N.G., 1991. The case against coarse-graining. Physica A 173 (3), 323–341.

Wallis, G.B., 1969. One-Dimensional Two-Phase Flow. McGraw-Hill Book Company, New York.

Yasseri, S., Rahnema, F., 2014. Consistent spatial homogenization in transport theory. Nucl. Sci. Eng. 176 (3), 292–305.